\documentclass[preprintnumbers,article,amsmath,amssymb,floatfix,10pt,prd,superscriptaddress,nofootinbib, twocolumn]{revtex4-2}

\usepackage{hyperref}
\hypersetup{
  colorlinks=true,        
  linkcolor=blue,         
  citecolor=cyan,         
}

\def\be{\begin{equation}}
	\def\ee{\end{equation}}
\def\bea{\begin{eqnarray}}
	\def\eea{\end{eqnarray}}

\usepackage{graphicx}
\usepackage{dcolumn}
\usepackage{bm}
\usepackage{color}
\usepackage{enumitem}
\usepackage{amsmath}
\usepackage{amssymb}

\newcommand{\beq}{\begin{equation}}
\newcommand{\eeq}{\end{equation}}

\usepackage{bbm}
\usepackage{amsfonts}
\usepackage{mathrsfs}
\usepackage{latexsym}
\usepackage{epsfig}
\usepackage{epstopdf}
\usepackage{epstopdf}
\usepackage{graphicx}
\usepackage{amssymb}
\usepackage{amsmath}
\usepackage{dcolumn}
\usepackage{bm}
\usepackage{color}
\usepackage{comment}
\usepackage{xcolor}
\usepackage{orcidlink}

\begin{document}

\title{Equatorial Periodic Orbits and Gravitational Wave Signatures in Weyl Geometric Gravity Black Holes}

\author{Dhruba Jyoti Gogoi \orcidlink{0000-0002-4776-8506}}
\email{moloydhruba@yahoo.in}
\affiliation{Department of Physics, Madhabdev University, Narayanpur, Lakhimpur 784164, Assam, India}

\author{Himanshu Chaudhary}
\email{himanshu.chaudhary@stud.ubbcluj.ro} 
\affiliation{Faculty of Physics, Babeș-Bolyai University, Kogălniceanu Street, Cluj-Napoca, 400084, Romania}

\author{Tiberiu Harko}
\email{tiberiu.harko@aira.astro.ro}
\affiliation{Faculty of Physics, Babeș-Bolyai University, Kogălniceanu Street, Cluj-Napoca, 400084, Romania}
\affiliation{Astronomical Observatory, 19 Cire\c silor Street, Cluj-Napoca 400487, Romania}

\begin{abstract}
Weyl geometric gravity theory is a gravitational theory with Lagrangian density constructed from the square of the Weyl curvature scalar and the strength of the Weyl vector. The theory can be reformulated as a  scalar - vector - tensor theory in the Riemann geometry by introducing an auxiliary scalar field, in the presence of a non - minimal coupling between the Ricci scalar and the scalar field. Assuming that the Weyl vector has only a radial component, an exact spherically symmetric black hole type solution of the field equations in the vacuum can be obtained. The solution depends on three arbitrary integration constants. As compared to the case of the Schwarzschild black hole, the Weyl geometric gravity black hole solution contains two new terms, linear and quadratic, in the radial coordinate. 
It is the main goal of the present work to investigate astrophysical properties of the Weyl geometric black holes by considering the behavior of the timelike geodesics, which allows to  explore how the geometric nonmetricity influences the strong-field dynamics of massive test particles as well as the topological structure of the periodic orbits. The resulting gravitational wave emission in extreme mass-ratio inspiral (EMRI) systems is also considered. Our results show that the  presence of nonmetricity in Weyl geometric gravity leaves distinct and measurable signatures on both the orbital topology and the emitted gravitational waveforms of compact EMRI systems. These unique temporal and amplitude modulations provide a promising observational avenue for testing the presence of nonmetricity on cosmic scales.
\end{abstract}

\maketitle

\date{\today}

\tableofcontents

\section{Introduction}

The general theory of relativity is one of the cornerstones  of modern physics and cosmology. In its initial formulation by Einstein \cite{Ein}  and Hilbert \cite{Hilb} general relativity was built up by using the mathematical formalism of the Riemannian geometry, which allows to  endow space-time with a metric, and an affine structure. The gravitational properties of the space-time follow from the geometry, and thus they characterized  by a metric, an affine connection, and by the Riemann curvature tensor, and its contractions.

Almost immediately after the birth of general relativity, Weyl proposed a generalization of Riemannian geometry \cite{Weyl1, Weyl2}, with the aim of developing a unified theory of electromagnetism and gravity. In Weyl's unified theory, the existence of the electromagnetic field is due to the presence of the nonmetricity of the space-time, a concept which does not exist in Riemann's geometry. A few years later, Cartan developed another generalization of the Riemann geometry, by introducing the torsion tensor, which describes a geometry with non-symmetric Christoffel symbols. The concept of torsion is at the foundation of the Einstein-Cartan gravity theory (see \cite{EC} for a review), in which torsion is generated by the spin of particles.  

 Weyl's geometry was generally ignored by physicists in its first 50 years of existence, mainly as a result of Einstein's criticism of the  interpretation of the nonmetricity from a physical point of view. But especially recently the interest of physicists in Weyl geometry did increase significantly, since it leads to the description from a geometric perspective of important physical and mathematical effects on both macroscopic and microscopic scales \cite{Scholz, Scholz1}. Weyl also postulated that all laws of physics must be satisfy the principle of invariance with respect to conformal transformations, which implies that they must remain unchanged under local conformal transformations of the form  $d\tilde{s}^2 = \Phi ^n(x)ds^2 = \Phi ^n(x)g_{\mu \nu}dx^\mu dx^\nu =  \tilde{g}_{\mu \nu}dx^\mu dx^\nu$. In a conformal transformation of the metric $\Phi (x)$ is called the conformal factor, and $n$ is the Weyl charge. The conformal transformations describe the variations of the units of length and time at each point of the space-time manifold. 

Several recent theoretical investigations have convincingly proved the important result that gravitation can be described by using three mathematically
equivalent geometric formalisms. These independent formalisms are based on the three basic geometric concepts of curvature, torsion, and non-metricity \cite{Jim}. The formulation of general relativity by using only non-metricity was proposed in \cite{Nester}, leading to the elaboration of the symmetric teleparallel theory of gravity. This direction of research  was generalized and it led to the formulation of $f(Q)$ gravity theory \cite{Jim1}, or coincident general relativity, and in which $f$ is an arbitrary function of its argument $Q$, representing the quadratic nonmetricity scalar.The $f(Q)$ theory is based on the fundamental action $S=\int{f(Q)\sqrt{-g}d^4x}$. The $f(Q)$ gravity theory was generalized by introducing a coupling between matter with nonmetricity \cite{Harko1, Harko2}, leading to the $f(Q,T$) theory \cite{Harko2}, in which the gravitational Lagrangian is given by an arbitrary  function $f$ of the trace $T$ of the matter-energy-momentum tensor and of the non-metricity $Q$.

The modified theories of gravity have recently become fundamental for the understanding of the cosmological and gravitational dynamics. This significant increase for modified gravity theories  was determined by the recent observational results that confronts general relativity with important theoretical challenges. These challenges come from several sets of observational data, related to dark matter and dark energy, and which cannot be explained using Einstein's general relativity without introducing at least one extra parameter in the theory. From the major challenges that present day cosmology must face  we mention the cosmic acceleration problem  \cite{46} and the Hubble tension \cite{47}.

In spite of its negative assessment by Einstein, Weyl's geometry was considered in many physical applications. By introducing a real scalar field $\phi$ of weight $w(\phi)=-1$, Dirac \cite{Di1,Di2} generalized Weyl's theory. To describe the gravitational interaction,   the Lagrangian density $L=-\phi^2R +kD_\mu \phi D^\mu \phi+c\phi^4 + (1/4)F_{\mu \nu}F^{\mu \nu}$ was introduced, where $R$ is the Ricci scalar, and $c$ and $k = 6$ are constants, and $F_{\mu \nu}$ we is the electromagnetic type tensor constructed from the Weyl vector. The Dirac Lagrangian is conformally invariant by construction. 

In a class of theories called conformal gravity \cite{Ma1,Ma2}, the principle of conformal invariance was applied via the action $I_W=-\alpha _g\int{C_{\lambda \mu \nu \sigma}C^{\lambda \mu \nu \sigma}\sqrt{-g}d^4x}$, where $\alpha_g$ is a coupling constant,  and $C_{\lambda \mu \nu \sigma}$ is the Weyl conformal tensor. The main advantage of conformal gravity is that it can solve the dark matter problem from a purely geometric perspective, without  the need of introducing new forms of matter. 

A new theoretical approach to Weyl conformally invariant gravity, based on the simplest action $S_W=\int{\tilde{R}^2\sqrt{-g}d^4x}$, where $\tilde{R}$ is the Weyl scalar,  was introduced in \cite{Gh1,Gh2,Gh3,Gh4,Gh5,Gh6,Gh7,Gh8,Gh9}. The corresponding theory is called the Weyl geometric gravity (WGG) theory, and it is based on the linearization of $S_W$ through the introduction of an auxiliary scalar field $\phi$. No additional degrees of freedom are introduced in the theory. Moreover,  through a geometric Stueckelberg mechanism the Weyl gauge symmetry $D(1)$ is broken spontaneously, leading to an Einstein-Proca type action. In this action the Weyl gauge field $\omega _\mu$ is naturally appears. A mimetic extension of the Weyl geometric gravity theory was proposed in \cite{Ha0}.   

Various cosmological and astrophysical implications of the WGG theory have also been considered recently. The cosmological implication of the theory have been investigated in \cite{Ha1,Ha2,Ha3}.  In particular, the coupling of matter and geometry in conformal quadratic Weyl gravity was investigated in \cite{Ha1}, by assuming a coupling term of the form $L_m\tilde{R}^2$, where $L_m$ is the matter Lagrangian, and $\tilde{R}$ is the Weyl scalar. The cosmological applications of the theory were also investigated in detail. The astrophysical properties of neutron, quark, and Bose-Einstein condensate stars in the  WGG theory were considered in \cite{Ha4}.  The analysis of the stellar structure performed for several equations of state of high density matter did show that in the WGG theory stars are more massive than in the standard general relativistic models. 

The existence of black holes is one of its most important predictions of general relativity is the existence. Two basic observations have provided convincing evidence for the existence of black holes. The first evidence,  providing direct evidence for the existence of black holes, is the detection by the LIGO and VIRGO experiments of the gravitational waves \cite{grav1,grav2}. Direct visual confirmation of the existence of black holes was provided by the Event Horizon Telescope (EHT) collaboration, which  obtained the first image showing the plasma orbiting the black hole at the center of galaxy M87 \cite{grav3}. An image of Sagittarius A*,  the supermassive black hole at the center of the Milky Way galaxy was also obtained by EHT \cite{grav4}. These observations confirmed the important predictions of general relativity, and they have also provided some  new perspectives for the  understanding of the structure and properties of compact astrophysical objects. 

In the framework of the WGG theory exact and numerical black hole solutions were obtained in \cite{Ha5}. The possibility of explaining dark matter as a Weyl geometric effect was considered in \cite{Ha6}.   A detailed analysis of the galactic rotation curves in the WGG theory, based on the exact black hole solution, was performed in \cite{Ha7}. The WGG gravity theoretical predictions were tested by using 175 galaxies from the Spitzer Photometry \& Accurate Rotation Curves (SPARC) database.  The exact solution of the Weyl geometric gravity can successfully describe a the large variety of the rotation curves of the SPARC sample, and thus give a satisfactory description of the particle motion in the galactic halos, without the need of introducing an extra dark matter component. 
  
The propagation of the electromagnetic wave packets, and the gravitational spin Hall effect of light, were investigated in the framework of the WGG theory in \cite{Ha8}. In  \cite{Ha9} the thermodynamic properties of the exact WGG black hole solution were investigated in \cite{Ha9}. The WGG  black holes have thermodynamic properties that strongly differentiate them from the standard solutions of general relativity and of the other extended gravity theories. These differences offer the possibility of a better understanding of the thermodynamic implications of modified gravity in black hole physics. The lensing properties in the strong field regimes of the WGG black hole were considered in \cite{Ha10}, and the shadow of the black hole was investigated, leading to a first set of constraints on the solution parameters by using the observational data from the shadows of the M87* and Sgr A* supermassive black holes. Several observational tests of the WGG black hole were considered in \cite{Ha11}.  

The investigation of strong-field gravitational environments has entered a transformative phase, catalyzed by the detection of gravitational waves and high-resolution imaging of supermassive black holes. Among the diverse probes used to scrutinize these regimes, periodic timelike geodesics provide uniquely informative data. In relativistic potentials, bound motion is organized into periodic orbits whenever the ratio of azimuthal to radial frequencies is rational. These trajectories constitute a natural ``skeleton" for bound motion, facilitating the transition from standard relativistic precession to the highly non-trivial zoom-whirl regime \cite{Grossman:2008yk,Levin:2008mq,Glampedakis:2002ya,Misra:2010pu,Levin:2009sk}.

The taxonomy of zoom-whirl orbits, defined by the integers $(z,w,v)$, has been utilized extensively to map orbital structures in various black hole spacetimes and alternative gravity scenarios \cite{Levin:2008ci,Fujita:2009bp,Healy:2009zm,Wei:2019zdf,Azreg-Ainou:2020bfl,Deng:2020yfm,Deng:2020hxw,Wang:2022tfo,QiQi:2024dwc,Alloqulov:2025ucf,Haroon:2025rzx,Wang:2025hla,Lu:2025cxx,Tu:2023xab,Yang:2024lmj,Jiang:2024cpe,Huang:2025vpi}. In these frameworks, orbital morphology is intrinsically tied to the effective potential, which is determined by the underlying spacetime geometry. Consequently, modifications to the metric functions—such as those found in different modified gravity theories, like $f(Q)$ gravity or symmergent models—alter the potential barrier and the resulting geodesic paths.

Such dynamics are particularly relevant for extreme mass-ratio inspirals (EMRIs), where a stellar-mass object orbits a supermassive black hole. During the adiabatic regime, this process is viewed as a sequence of evolving bound geodesics. Since the gravitational radiation emitted during this phase depends on orbital frequencies and near-horizon motion, modifications to the background geometry leave distinct imprints on the waveform. This makes periodic orbits an essential theoretical bridge between geodesic dynamics and observable gravitational wave features.

The Weyl geometric gravity theory admits in static spherical symmetry an exact solution of the vacuum gravitational field equations \cite{Ha5,Ha6}-the WGG black hole. The WGG solution generalizes  the Schwarzschild - de Sitter solution of standard general relativity, through the inclusion of two new terms in the metric, as compared to the Schwarzschild case. These new terms have a linear and a quadratic dependence on the radial coordinate, respectively. The WGG solution is derived under the assumption that the Weyl vector $\omega _\mu$ has only a radial component, so that $\omega _\mu =\left(0,\omega _1,0,0\right)$. The WGG solution is three-parametric, and it depends on three arbitrary integration constants. One of these constants can be fixed From physical considerations as representing the mass of the central objecttwo of these constants can be fixed as the gravitational radius (mass) of the central compact object. Therefore, the solution depends on two free parameters,  which appear in the metric as a constant and independent on $r$ term, which also determines the linearly increasing term, and a second constant describing the $r^2$ term.

Motivated by these theoretical advancements, the purpose of this work is to investigate equatorial periodic orbits and their gravitational wave signatures in the context of Weyl geometric gravity black holes. By analyzing the effective potential and orbital ``skeleton," we aim to determine how Weyl geometric parameters influence particle dynamics and detectable signals.

The present paper is organized as follows. The foundations of the Weyl geometry and the fundamental description of the Weyl geometric gravity theory are presented in Section~\ref{sect1}. The metric of the exact solution of the vacuum field equations is also written down. The behavior of the orbits of the massive particles in the WGG black hole geometry is analyzed in Section~\ref{sect2}. The gravitational signals emitted by particle gravitating in circular orbits around the black hole is investigated in Section~\ref{sect3}. Finally, we discuss and conclude our results in Section~\ref{sect4}.

\section{Black Hole Solutions in the Weyl Geometric Gravity theory}\label{sect1}

In the present Section we will briefly review the foundations of the Weyl geometry, and we introduce the action and field equations of Weyl geometric gravity. For the action we adopt its simplest form, as inspired by Weyl-geometry, and which is formulated the satisfy the requirement of the conformal invariance of the physical laws. The action is quadratic in the Weyl scalar, and we linearize it by introducing an auxiliary scalar field. Then we reformulate the gravitational action in Riemann geometry. Moreover, we present the exact solution of the vacuum field equations of the WGG theory that generalizes the Schwarzschild-de Sitter solution.

 \subsection{Fundamentals of Weyl geometry}

A few years after Einstein and Hilbert (by using a variational principle) derived the field equations of general relativity,  Weyl \cite{Weyl1, Weyl2} extended Riemann geometry by postulating the existence in every point $x$ of the space-time of a conformal transformation. A conformal transformation $g_{\mu \nu}\rightarrow \tilde{g}_{\mu \nu}$ for the  metric tensor $g_{\mu \nu}$  is defined according to
\be
\tilde{g}_{\mu \nu } =\Sigma ^{n}(x)g_{\mu \nu },
\ee
where by $\Sigma (x)$ we have denoted  an arbitrary function, and $n$ is the Weyl charge. In the following we consider only the case $n=1$. In Weyl geometry the length $l$ of a vector varies when it is parallely transported along an infinitesimal path,  from the point $x^\mu$ to the point $x^\mu + \delta
x^\mu$. The change in the length of a vector is given by
\begin{eqnarray}  \label{a1}
\delta l = l \omega_\mu \delta x^\mu,
\end{eqnarray}
where $\omega_\mu$ is the Weyl vector field. The field $\omega_\mu$ is the source of the nonmetricity $Q_{\lambda \mu \nu}$ of the Weyl geometry, which leads   to the nonconservation of the covariant derivative of the metric tensor, which in the Weyl geometry is defined as
\begin{align}\label{nmetr}
& \tilde{\nabla}_\lambda g_{\mu\nu} = - \alpha \omega_\lambda g_{\mu\nu}=Q_{\lambda \mu \nu},
\end{align}
where $\alpha$ is called the Weyl gauge coupling constant. From Eq.~(\ref{nmetr}) the connection of the Weyl geometry is obtained as
\begin{align}\label{connex}
& \tilde{\Gamma}^\lambda_{\mu\nu} = \Gamma^\lambda_{\mu\nu} + \frac{1}{2}
\alpha \Big[ \delta^\lambda_\mu \omega_\nu + \delta^\lambda_\nu \omega_\mu -
g_{\mu\nu} \omega^\lambda \Big],
\end{align}
and it represents a generalization of the Levi-Civita connection of the Riemann geometry. Wee denote in the following by a tilde the physical and geometrical quantities defined on a Weyl manifold.

The curvature tensor of the Weyl geometry is defined as
\begin{equation}
\tilde{R}_{\mu \nu \sigma }^{\lambda }=\partial _{\nu }\tilde{\Gamma}_{\mu
\sigma }^{\lambda }-\partial _{\sigma }\tilde{\Gamma}_{\mu \nu }^{\lambda }+%
\tilde{\Gamma}_{\rho \nu }^{\lambda }\tilde{\Gamma}_{\mu \sigma }^{\rho }-%
\tilde{\Gamma}_{\rho \sigma }^{\lambda }\tilde{\Gamma}_{\mu \nu }^{\rho },
\end{equation}%
giving by first and second contraction
\begin{equation}
\tilde{R}_{\mu \nu }=\tilde{R}_{\mu \lambda \nu }^{\lambda },\tilde{R}%
=g^{\mu \sigma }\tilde{R}_{\mu \sigma },
\end{equation}%
respectively. The Weyl scalar $\tilde R$ is obtained as
\begin{equation}\label{R}
\tilde{R}=R-3n\alpha \nabla _{\mu }\omega ^{\mu }-\frac{3}{2}\left( n\alpha
\right) ^{2}\omega _{\mu }\omega ^{\mu },
\end{equation}
where $R$ denotes the Ricci scalar, defined in the Riemann geometry.

Another geometric quantity, the Weyl tensor, which plays a central role in the conformally invariant gravitational theories \cite{Ma1,Ma2}, is defined according to
\begin{equation}
\tilde{C}_{\mu \nu \rho \sigma }^{2}=C_{\mu \nu \rho \sigma }^{2}+\frac{3}{2}%
\left( \alpha n\right) ^{2}\tilde{F}_{\mu \nu }^{2},
\end{equation}
where $C_{\mu \nu \rho \sigma }$ is the Weyl tensor, defined in the standard Riemannian geometry \cite{Ma1,Ma2}, and 
\be
 \tilde{F}_{\alpha\beta} =  \tilde{\nabla}_{[\alpha} \omega_{\beta]} =  \nabla_{[\alpha}
\omega_{\beta]} =  \partial_{[\alpha} \omega_{\beta]}=\partial _{\alpha}\omega _\beta-\partial _\beta \omega _\alpha,
\ee
is the Weyl field strength.

After a conformal transformation with a conformal factor $\Sigma$ in a point $x$, the transformation laws of the
metric tensor, of the Weyl field, and of an arbitrary  scalar field $\phi $ are obtained as
\begin{eqnarray}  \label{a2}
\hat g_{\mu\nu}(x) &=& \Sigma (x)g_{\mu\nu}(x), \hat \omega _\mu (x)= \omega_\mu (x)-
\frac{1}{\alpha} \partial_\mu \ln\Sigma (x), \nonumber\\
\hat \phi (x)&=& \Sigma^{-\frac{1}{2}} (x)\phi (x).
\end{eqnarray}

In the following,  we assume that the Weyl vector field is a purely geometric quantity. As for $\omega _\mu$, we interpret it as a gauge field mediating at different space-time points the conformal factors \cite{Gh1,Gh2}.

\subsection{Action and field equations of Weyl geometric gravity}

Weyl geometric gravity is constructed in Weyl geometry, and it is obtained from the simplest conformally invariant Lagrangian density,  given by \cite{Weyl1, Weyl2, Gh1,Gh2}
\bea\label{inA}
L_0=\Big[\, \frac{1}{4!}\,\frac{1}{\xi^2}\,\tilde R^2  - \frac14\, F_{\mu\nu}^{\,2} \Big]\sqrt{-g}.
\eea
The perturbative coupling constant $\xi$ satisfies the condition  $\xi < 1$. In the action (\ref{inA}) a scalar degree of freedom is hidden, which can be obtained explicitly by replacing the term  $\tilde{R}^2$ by
$\tilde{R}^2\rightarrow -2 \phi_0^2\,\tilde R-\phi_0^4$, where $\phi_0$ is the hidden (auxiliary)  scalar field. The Lagrangian density thus obtained  by  is equivalent with the initial one, since
after replacing in the new Lagrangian
$L_1$ the solution $\phi_0^2=-\tilde R$ of the equation of motion of $\phi_0$, we recover the initial action  Eq.~(\ref{inA}).

Hence, after linearizing the quadratic term in the action (\ref{inA}), we obtain an equivalent Lagrangian which we call the Weyl geometric gravity Lagrangian, and which is given by
\bea\label{alt3}
L_1=\sqrt{-g} \Big[-\frac{1}{12}\frac{1}{\xi^2}\,\phi_0^2\,\tilde R
-\frac14 \,F_{\mu\nu}^2-\frac{\phi_0^4}{4!\,\xi^2}
\Big].
\eea

The Weyl geometric gravity Lagrangian density (\ref{alt3}) is the simplest possible Lagrangian density that fully implements conformal invariance and  Weyl gauge symmetry. Also $L_1$  spontaneously breaks down to an Einstein-Proca type Lagrangian of the Weyl gauge field.

By replacing in Eq.~(\ref{alt3})  $\tilde{R}$ by its Riemannian counterpart, as given by Eq.~(\ref{R}), we obtain the Weyl geometric gravity action, defined in the Riemann space,  given by \cite{Gh3,Gh4,Gh5},
\begin{eqnarray}\label{a3}
\mathcal{S }&=& \int d^4x \sqrt{-g} \Bigg[ -\frac{1}{12} \frac{\phi^2}{\xi^2}
\Big( R - 3\alpha\nabla_\mu \omega^\mu - \frac{3}{2} \alpha^2 \omega_\mu
\omega^\mu \Big) \nonumber\\
&&- \frac{1}{4!}\frac{\phi^4}{\xi^2} - \frac{1}{4} F_{\mu\nu}
F^{\mu\nu} \Bigg].
\end{eqnarray}

The theory described by the above action is called the Weyl geometric gravity theory.  After varying the action Eq.~(\ref{a3}) with respect to the metric tensor we obtain the gravitational field
equations of the Weyl geometric gravity theory as \cite{Ha1,Ha2},
\begin{eqnarray}\label{b2a}
&&\frac{\phi ^{2}}{\xi ^{2}}\Big(R_{\mu \nu }-\frac{1}{2}Rg_{\mu \nu }%
\Big)-\frac{1}{4\xi ^{2}}\phi ^{4}g_{\mu \nu }\nonumber\\
&&-\frac{3\alpha }{2\xi ^{2}}\Big(\omega ^{\rho }\nabla _{\rho }\phi
^{2}g_{\mu \nu }-\omega _{\nu }\nabla _{\mu }\phi ^{2}-\omega _{\mu }\nabla
_{\nu }\phi ^{2}\Big)\nonumber\\
&&+\frac{3\alpha ^{2}}{4\xi ^{2}}\phi ^{2}\Big(\omega
_{\rho }\omega ^{\rho }g_{\mu \nu }-2\omega _{\mu }\omega _{\nu }\Big)
+6F_{\rho \mu }F_{\sigma \nu }g^{\rho \sigma }-\frac{3}{2}F_{\rho \sigma
}^{2}g_{\mu \nu }\nonumber\\
&&
+\frac{1}{\xi ^{2}}%
\Big(g_{\mu \nu }\Box -\nabla _{\mu }\nabla _{\nu }\Big)\phi ^{2}=0.
\end{eqnarray}

The trace of Eq.~(\ref{b2a}) gives the relation
\begin{equation}\label{b3n}
\Phi R+3\alpha \omega ^{\rho }\nabla _{\rho }\Phi +\Phi ^{2}-\frac{3}{2}%
\alpha ^{2}\Phi \omega _{\rho }\omega ^{\rho }-3\Box \Phi =0,
\end{equation}%
where we have denoted $\Phi \equiv \phi ^{2}$.

By varying the action  Eq.~(\ref{a3}) with respect to the scalar field $\phi $ we obtain the equation of motion of $%
\Phi $,
\begin{equation}\label{b4}
R-3\alpha \nabla _{\rho }\omega ^{\rho }-\frac{3}{2}\alpha ^{2}\omega _{\rho
}\omega ^{\rho }+\Phi =0.
\end{equation}%

From  Eqs.~(\ref{b4}) and (\ref{b3n}) the generalized Klein-Gordon equation for the scalar field $\Phi$ is found as
\begin{equation}\label{b5}
\Box \Phi-\alpha \nabla _{\rho }(\Phi\omega ^{\rho })=0.
\end{equation}%

The equation of motion of the Weyl vector is given by
\begin{equation}\label{Fmunu}
4\xi ^{2}\nabla _{\nu }F^{\mu \nu }-\alpha ^{2}\Phi\omega ^{\mu }+\alpha \nabla
^{\mu }\Phi=0.
\end{equation}%

After applying to both sides of Eq.~(\ref{Fmunu}) the operator  $\nabla _{\mu }$,  Eq.~(\ref{b5}) is reobtained again. This result shows the mathematical consistency of the Weyl geometric gravity  theory.

\subsection{The Weyl geometric black hole}\label{subsect_wgg_bh}

In the following we adopt the natural system of units, with $c=G=1$. In the static spherically symmetric geometry the space-time interval can be represented as
\begin{equation}\label{line}
ds^{2}=e^{\nu (r)}dt^{2}-e^{\lambda (r)}dr^{2}-r^{2}d\Omega ^{2},
\end{equation}%
where $d\Omega ^{2}=d\theta ^{2}+\sin ^{2}\theta d\varphi ^{2}$, and we have introduced a system of coordinates  $\left(t,r,\theta, \varphi\right)$, where $t$ is the time coordinate, $r$ is the radial coordinate, while $\theta$ and $\varphi$ are the angular coordinates. 

Due to the choice of the metric, in a spherically symmetric gravitational field in Weyl geometry,  the third and the
fourth components of the Weyl vector  must vanish, and hence the Weyl
vector has the components $\omega _{\mu }=\left( \omega _{0},\omega
_{1},0,0\right) $ \cite{Ha5,Ha6}.
In the following we consider solutions of the field equations of the Weyl geometric gravity theory only for the case in which the temporal component, $\omega _{0}$,
of the Weyl vector vanishes identically. Hence, for the Weyl vector $\omega _{\mu}$ we assume  the form
\be
\omega _{\mu }=(0,\omega _{1},0,0),
\ee

The system of the field equations (\ref{b2a})-(\ref{Fmunu}) of the Weyl geometric gravity theory does admit an exact vacuum solution in spherical symmetry. To obtain it it is assumed that the metric tensor components $\nu (r)$ and $\lambda (r)$ satisfy the mathematical condition
\be
\nu (r)+\lambda (r)=0,
\ee
assumed to be valid for all $r>0$.

Then, from the field equations we find \cite{Ha5,Ha6}
\be\label{40}
\Phi(r)=\frac{C_1}{\left(r+C_2\right)^2},
\ee
where $C_1$ and $C_2$ are arbitrary integration constants that must be determined from the boundary or initial  conditions, and 
\be
\omega _1=\frac{1}{\alpha}\frac{\Phi '}{\Phi}=-\frac{2}{\alpha}\frac{1}{r+C_2},
\ee
with the property $\lim_{r\rightarrow \infty}\omega _1(r)=0$. In the limit of large $r$ we have 
\be
\lim_{r\rightarrow \infty}\Phi(r)=0,
\ee
a condition showing that $\Phi$ vanishes at infinity.
 
Finally, the metric of the Weyl geometric gravity black hole is obtained as \cite{Ha5,Ha6}
\begin{equation}\label{ds}
 d s^2=-f(r) d t^2+\dfrac{1}{f(r)} d r^2+r^2\left(d\theta^2+\sin^2\theta d\phi^2\right),
 \end{equation}
with
\begin{equation}
    f(r)= \alpha -\frac{2 M}{r}-\frac{\left(\alpha ^2-1\right) }{6 M}r +\frac{c_3 }{4 M^2}r^2,
\end{equation}
where $M$ is the mass of the central object, and $\alpha $ and $c_3$ are constants. 

\section{Particle Orbits in Weyl Geometric Black Holes}\label{sect2}

In this section, we investigate the motion of test particles in the background of an exact black hole solution obtained in Weyl geometric gravity theory. The dynamics of particle motion in this static and spherically symmetric geometry  is governed by the geodesic equations, which can be derived from the Lagrangian of a free particle in curved spacetime,
\begin{equation}
\mathcal{L}=\frac{1}{2} g_{\mu\nu}\dot{x}^{\mu}\dot{x}^{\nu}=\delta/2,
\label{lagrangian_weyl}
\end{equation}
where the overdot denotes differentiation with respect to the affine parameter $\tau$, and $\delta=-1$ and $0$ correspond to massive and massless particles, respectively. Due to spherical symmetry, we restrict motion to the equatorial plane by setting $\theta=\pi/2$ and $\dot{\theta}=0$.

The generalized four-momentum associated with the Lagrangian \eqref{lagrangian_weyl} ensures that since the metric is independent of $t$ and $\phi$, the corresponding momenta are conserved:
\begin{gather}
p_t=g_{tt}\dot{t}=-E, \label{pt} \
p_\phi=g_{\phi\phi}\dot{\phi}=L, \
p_r=g_{rr}\dot{r}, \
p_\theta=0,
\end{gather}
where $E$ and $L$ represent the conserved specific energy and specific angular momentum per unit mass, respectively. Using the metric \eqref{ds}, the first integrals of motion become:
\begin{gather}
\dot{t}=\frac{E}{f(r)}, \
\dot{\phi}=\frac{L}{r^2}, \
\dot{r}^2=E^2 - f(r)\left(-\delta+\frac{L^2}{r^2}\right)
\label{rdot_weyl}
\end{gather}

For timelike geodesics ($\delta=-1$), the radial equation can be written in the effective potential form,
\begin{equation}
\dot{r}^2+V_{\mathrm{eff}}(r)=E^2,
\label{rdoteff_weyl}
\end{equation}
with the effective potential defined as:
\begin{equation}
V_{\mathrm{eff}}(r)=f(r)\left(1+\frac{L^2}{r^2}\right).
\label{veff_weyl}
\end{equation}

The effective potential encodes the combined influence of the nonmetricity of Weyl geometry on particle motion. The allowed regions of motion are determined by $E^2\geq V_{\mathrm{eff}}(r)$, while circular and bound orbits correspond to extrema of $V_{\mathrm{eff}}(r)$.

\begin{figure*}[htbp]
\centerline{
\includegraphics[scale = 0.58]{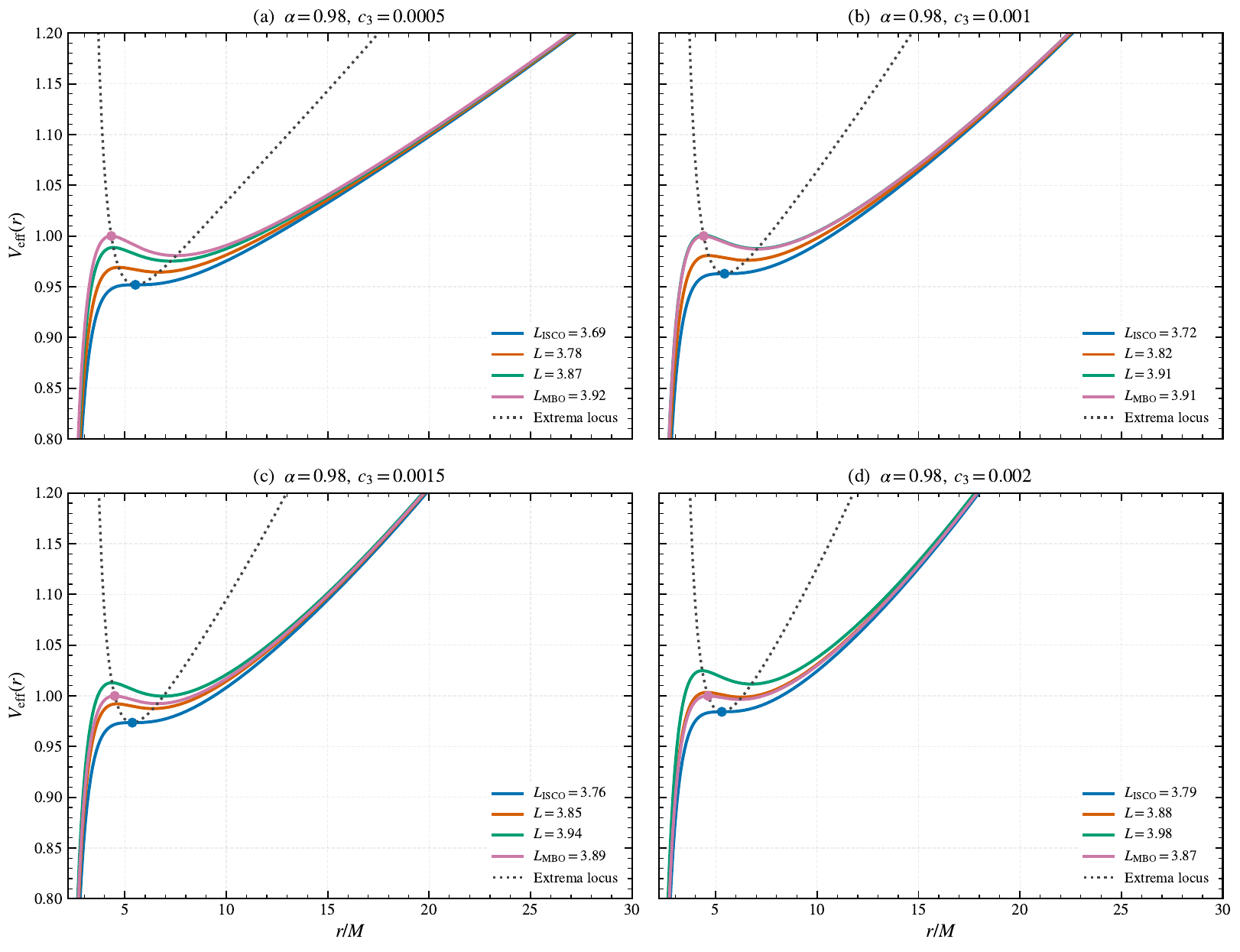}}
\caption{Effective potential $V_{\mathrm{eff}}(r)$ of the Weyl geometry black hole for varying $c_3$. The curves correspond to representative values of the angular momentum, including $L_{\mathrm{ISCO}}$ and $L_{\mathrm{MBO}}$. The dotted line denotes the locus of extrema of the potential.}
\label{fig01}
\end{figure*}

\begin{figure*}[htbp]
\centerline{
\includegraphics[scale = 0.58]{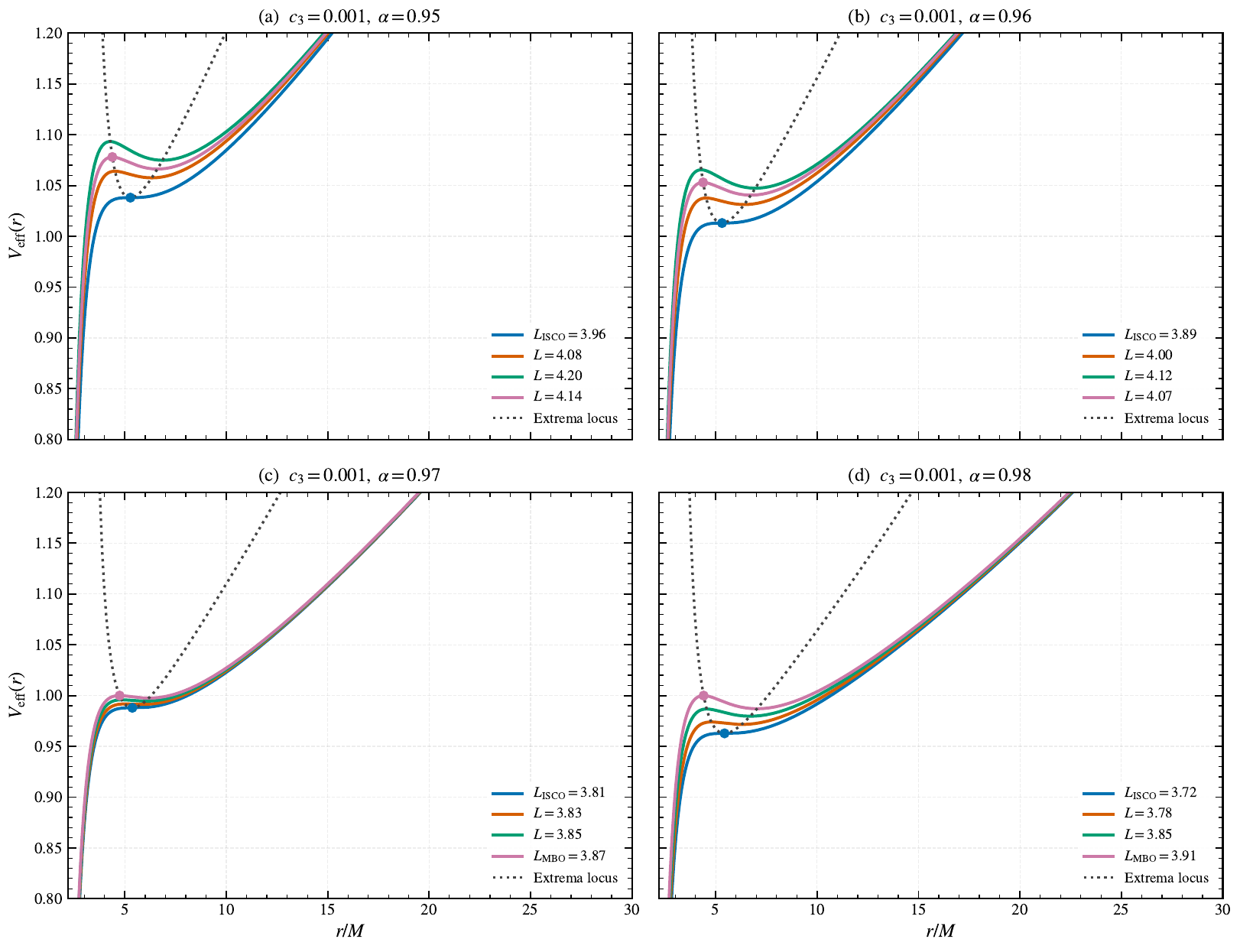}}
\caption{Effective potential $V_{\mathrm{eff}}(r)$ of the Weyl geometry black hole for varying $\alpha$. The curves correspond to representative values of the angular momentum, including $L_{\mathrm{ISCO}}$ and $L_{\mathrm{MBO}}$. The dotted line denotes the locus of extrema of the potential.}
\label{fig02}
\end{figure*}

Figures~\ref{fig01} and \ref{fig02} illustrate the effective potential $V_{\mathrm{eff}}(r)$ for timelike geodesics in the Weyl geometric spacetime for several values of the parameters $c_3$ and $\alpha$. Each panel shows representative curves corresponding to different values of the angular momentum, including the innermost stable circular orbit $L_{\mathrm{ISCO}}$ and the marginally bound orbit $L_{\mathrm{MBO}}$. The dotted line denotes the locus of extrema of the potential, which separates stable and unstable circular trajectories.

For a given parameter set, the effective potential exhibits a local minimum and a local maximum. The minimum corresponds to a stable circular orbit, while the maximum represents an unstable circular orbit. Bound motion is allowed in the region where $E^2$ lies between these two extrema, whereas plunging or unbound trajectories occur when the energy exceeds the potential barrier.

As the Weyl parameter $\alpha$ deviates from unity, the height of the potential barrier and the depth of the potential well are systematically modified. Physically, this indicates that the nonmetricity of the Weyl geometry alters the effective gravitational attraction experienced by the particle, leading to different binding energies for the orbits. Consequently, the locations of both the ISCO and MBO shift, and the corresponding angular momenta $L_{\mathrm{ISCO}}$ and $L_{\mathrm{MBO}}$ adjust monotonically with $\alpha$ and $c_3$.

Overall, Figures~\ref{fig01} and \ref{fig02} demonstrate that Weyl geometric gravity introduces a controlled deformation of the effective potential, directly influencing the orbital structure, accretion dynamics, and gravitational interactions of compact objects around Weyl black holes.

\subsection{Marginally Bound Orbits and Bound Motion Thresholds}\label{s4sub1}

In asymptotically flat spacetimes, a marginally bound orbit (MBO) represents the limiting unstable circular trajectory of a particle whose specific energy equals the rest-mass threshold at spatial infinity, $E^2_{\mathrm{MBO}} = 1$. The conditions defining such a critical orbit are given by
\begin{equation}
V_{\mathrm{eff}}(r)=E^2_{\mathrm{MBO}},\qquad
\frac{dV_{\mathrm{eff}}}{dr}=0,
\label{mbo_weyl}
\end{equation}
where $E_{\mathrm{MBO}}$ denotes the energy at the separatrix between bound and plunging or escaping trajectories. For any circular orbit at radius $r_c$ satisfying $V'_{\mathrm{eff}}(r_c)=0$, the specific angular momentum and energy take the form
\begin{equation}
L_c^2 = \frac{r_c^3 f'(r_c)}{2f(r_c)-r_c f'(r_c)}, \qquad
E_c^2 = \frac{2f(r_c)^2}{2f(r_c)-r_c f'(r_c)}.
\label{circ_EL_weyl}
\end{equation}

In the present Weyl geometric black hole spacetime, the metric function $f(r)$ contains linear and quadratic radial terms and is therefore not generally asymptotically flat. In particular, for $c_3 > 0$, the metric function grows quadratically at large $r$, causing $V_{\mathrm{eff}}(r)$ to diverge as $r \to \infty$ rather than approaching unity. Consequently, $E_c^2 = 1$ does not strictly represent an escape threshold to infinity. Nevertheless, in the regime of small Weyl deviations ($\alpha \approx 1$ and $|c_3| \ll 1$) pertinent to local strong-field dynamics, setting the formal reference threshold $E_{\mathrm{MBO}}^2 = 1$ in Eq.~\eqref{mbo_weyl} yields the inner unstable circular orbit condition
\begin{equation}
2f(r_{\mathrm{MBO}})^2 - 2f(r_{\mathrm{MBO}}) + r_{\mathrm{MBO}} f'(r_{\mathrm{MBO}}) = 0.
\label{rmbo_cond_weyl}
\end{equation}
Solving Eq.~\eqref{rmbo_cond_weyl} for $r_{\mathrm{MBO}}$ and substituting it into Eq.~\eqref{circ_EL_weyl} determines the corresponding reference angular momentum $L_{\mathrm{MBO}}$. More generally, for an arbitrary choice of $(\alpha, c_3)$ and angular momentum $L > L_{\mathrm{ISCO}}$, non-plunging bound motion is confined between the radial turning points where $E^2 = V_{\mathrm{eff}}(r; L)$, bounded from above by the local maximum of the inner potential barrier, $E^2_{\max}(L) = V_{\mathrm{eff}}(r_{\mathrm{un}}; L)$, and from below by the local minimum at the stable circular orbit, $E^2_{\min}(L) = V_{\mathrm{eff}}(r_{\mathrm{st}}; L)$. Both the barrier height and the resulting critical quantities $(r_{\mathrm{MBO}}, L_{\mathrm{MBO}})$ depend explicitly on $\alpha$ and $c_3$, demonstrating that geometric nonmetricity systematically shifts the strong-field capture threshold relative to the standard Schwarzschild limit ($r_{\mathrm{MBO}} = 4M$, $L_{\mathrm{MBO}} = 4M$).

\subsection{Innermost Stable Circular Orbits}\label{s4sub2}

The innermost stable circular orbit (ISCO) represents the smallest radius for which stable circular motion is possible. It is determined by the inflection point of the effective potential, satisfying the conditions:
\begin{equation}
V_{\mathrm{eff}}(r)=E^2,\qquad
\frac{dV_{\mathrm{eff}}}{dr}=0,\qquad
\frac{d^2V_{\mathrm{eff}}}{dr^2}=0.
\label{isco_weyl}
\end{equation}
From these equations, the ISCO angular momentum and energy are obtained as

\begin{equation}
    L_{\mathrm{ISCO}}^2=
\frac{r_{\mathrm{ISCO}}^3 f'(r_{\mathrm{ISCO}})}
{2f(r_{\mathrm{ISCO}})-r_{\mathrm{ISCO}} f'(r_{\mathrm{ISCO}})}, 
\end{equation}

\begin{gather}
E_{\mathrm{ISCO}}^2=
\frac{2f(r_{\mathrm{ISCO}})^2}
{2f(r_{\mathrm{ISCO}})-r_{\mathrm{ISCO}} f'(r_{\mathrm{ISCO}})},
\end{gather}
where the prime denotes differentiation with respect to $r$. The ISCO radius satisfies:
\begin{align}
r_{\mathrm{ISCO}} f(r_{\mathrm{ISCO}}) f''(r_{\mathrm{ISCO}}) - 2 r_{\mathrm{ISCO}} (f'(r_{\mathrm{ISCO}}))^2 \\ \notag + 3 f(r_{\mathrm{ISCO}}) f'(r_{\mathrm{ISCO}})=0.
\label{riscocond_weyl}
\end{align}

In the limit $\alpha\to 1$ and $c_3\to 0$, these expressions reduce to the standard Schwarzschild ISCO at $r_{\mathrm{ISCO}}=6M$. The presence of the Weyl parameters introduces deviations at both small and large radii.

\begin{figure*}[htbp]
\centerline{
\includegraphics[scale = 0.45]{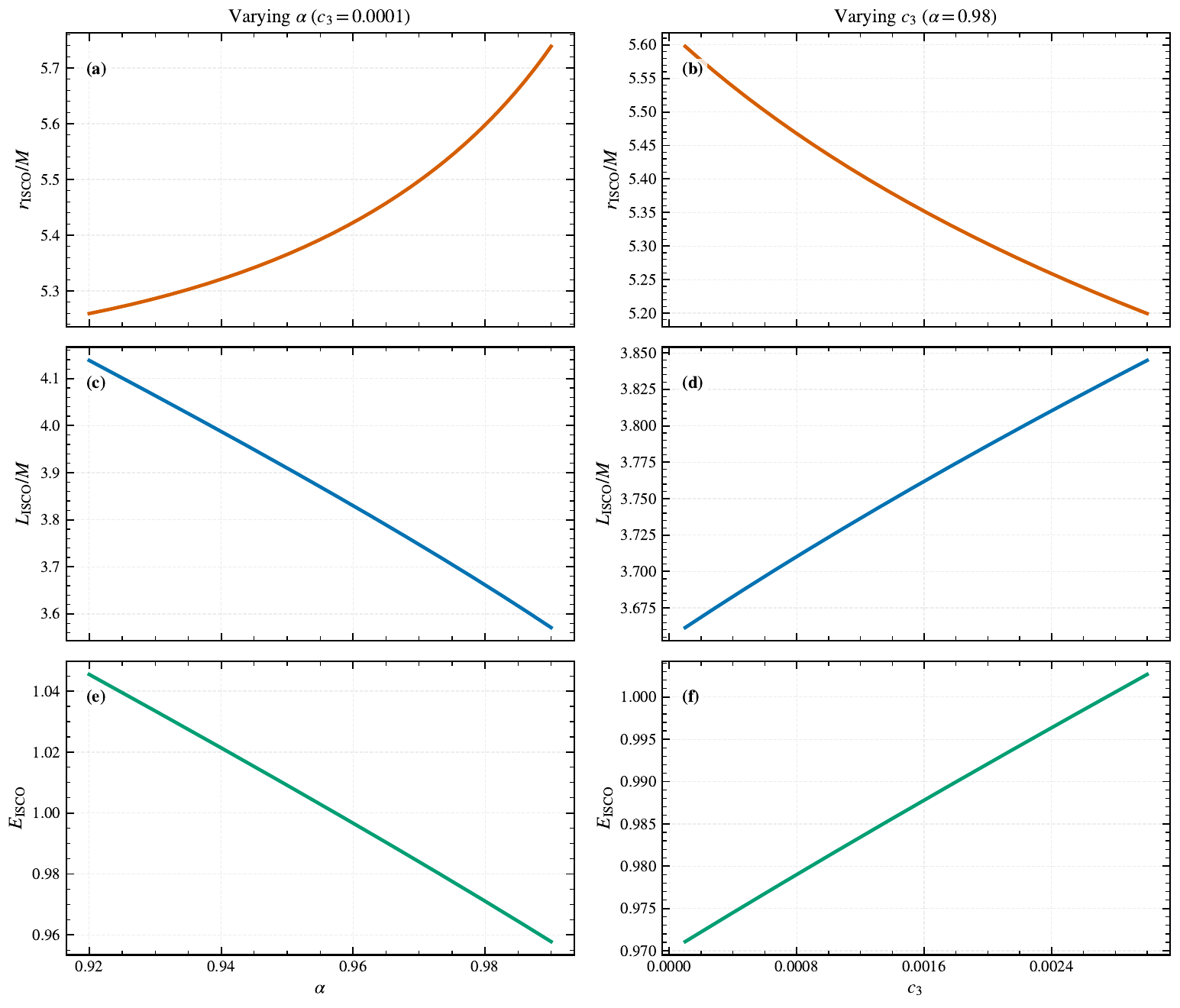}}
\caption{Dependence of the ISCO radius $r_{\mathrm{ISCO}}$, ISCO angular momentum $L_{\mathrm{ISCO}}$, and ISCO energy $E_{\mathrm{ISCO}}$ on the metric parameters $\alpha$ and $c_3$.}
\label{fig03}
\end{figure*}


Figure~\ref{fig03} displays the dependence of the ISCO radius $r_{\mathrm{ISCO}}$, the ISCO angular momentum $L_{\mathrm{ISCO}}$, and the ISCO energy $E_{\mathrm{ISCO}}$ on the Weyl geometric parameters $\alpha$ and $c_3$. As $\alpha$ increases, $r_{\mathrm{ISCO}}$ expands non-linearly, while $L_{\mathrm{ISCO}}$ and $E_{\mathrm{ISCO}}$ exhibit a decreasing trend. Conversely, an increase in $c_3$ pulls the ISCO inward while requiring higher angular momentum and energy to maintain stability. Physically, this reflects a redistribution of the effective potential induced by the nonmetricity tensor, which shifts the stability threshold dynamically.

\begin{figure}[htbp]
\centerline{
\includegraphics[scale = 0.45]{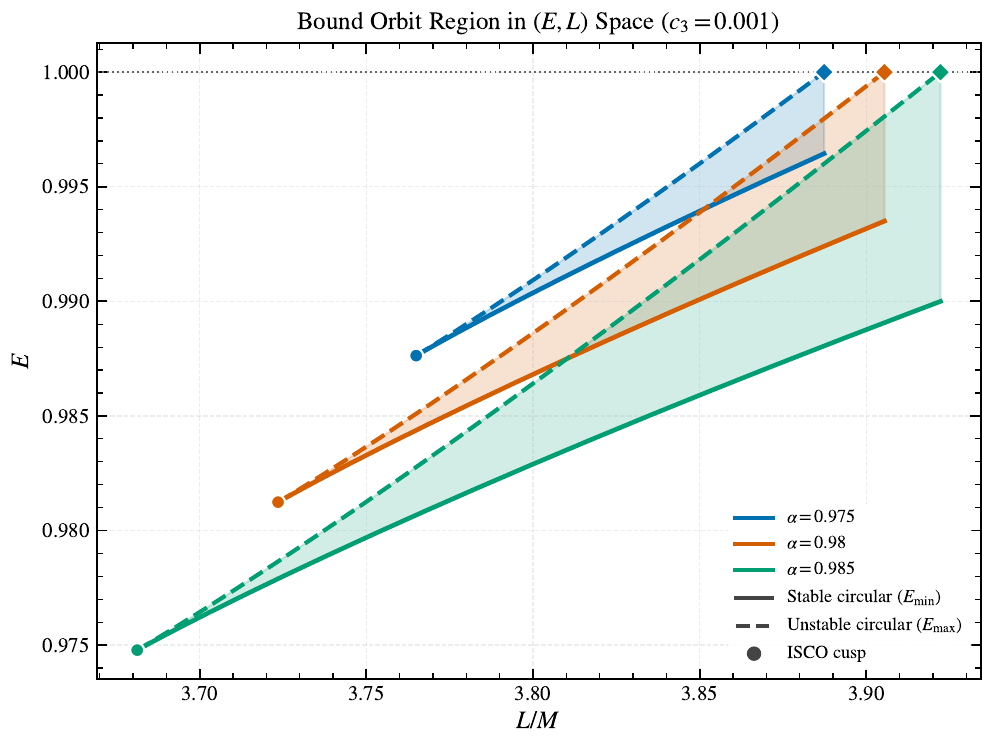}}
\caption{Allowed parameter regions for bound timelike orbits in the $E$--$L$ plane for different values of $\alpha$.}
\label{fig04}
\end{figure}

\begin{figure}[htbp]
\centerline{
\includegraphics[scale = 0.45]{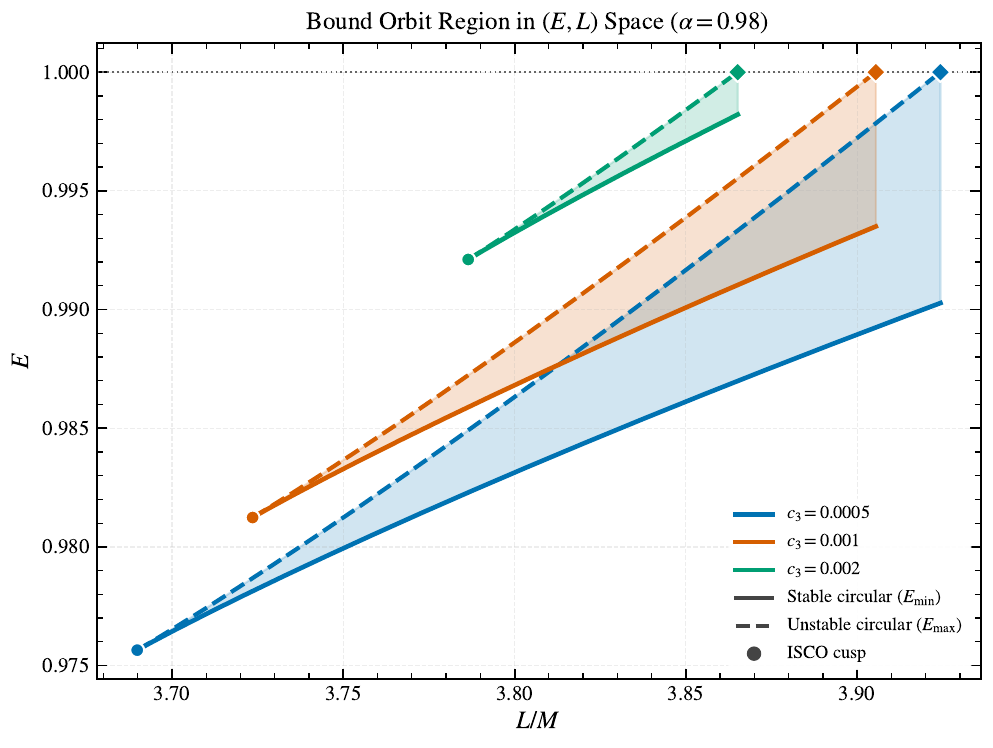}}
\caption{Allowed parameter regions for bound timelike orbits in the $E$--$L$ plane for different values of $c_3$.}
\label{fig05}
\end{figure}

Figures~\ref{fig04} and \ref{fig05} show the allowed parameter regions for bound timelike orbits in the $(E,L)$ plane. The shaded areas represent the domain in which bound motion is possible, bounded from below by the ISCO curve and from above by the marginally bound orbit limit. As the geometric parameters vary, the admissible region shrinks or expands, modifying the phase space available for long-lived bound orbits. This indicates that Weyl gravity facilitates plunging or escape orbits for a wider or narrower range of initial conditions depending on the precise values of $\alpha$ and $c_3$.

\subsection{Periodic Orbits and Rational Parameter}\label{s4sub3}

Periodic orbits correspond to bound trajectories for which the ratio of angular and radial frequencies is rational. The rational parameter $q$ is defined as
\begin{equation}
q=\frac{\omega_\phi}{\omega_r}-1
=\frac{\Delta\phi}{2\pi}-1,
\label{qdef_weyl}
\end{equation}
where $\Delta\phi$ is the total azimuthal angle accumulated during one complete radial oscillation between the periapsis $r_p$ and apoapsis $r_a$,
\begin{equation}
\Delta\phi
=2\int_{r_p}^{r_a}
\frac{L}{r^2\sqrt{E^2-f(r)\left(1+\frac{L^2}{r^2}\right)}}\,dr.
\label{deltaphi_weyl}
\end{equation}

The rational number $q$ can be written in terms of three integers $(z,w,v)$ \cite{Levin:2008mq},
\begin{equation}
q=w+\frac{v}{z},
\label{zwv_weyl}
\end{equation}
where $z$ is the zoom number, $w$ is the number of whirls, and $v$ is the vertex number. This zoom-whirl structure characterizes the topology of strong-field orbits around the Weyl geometric black hole.

For bound motion, the angular momentum is restricted to
\begin{equation}
L_{\mathrm{ISCO}}\leq L\leq L_{\mathrm{MBO}},
\end{equation}
and it is convenient to parametrize it as
\begin{equation}
L=L_{\mathrm{ISCO}}
+\epsilon\left(L_{\mathrm{MBO}}-L_{\mathrm{ISCO}}\right),
\qquad 0<\epsilon<1.
\label{epsilon_weyl}
\end{equation}
By fixing $\epsilon$ and solving Eq.~\eqref{qdef_weyl} for the energy $E$, one can systematically construct families of periodic orbits and analyze how Weyl nonmetricity modifies their geometric structure.

\begin{figure}[htbp]
\centerline{
\includegraphics[scale = 0.45]{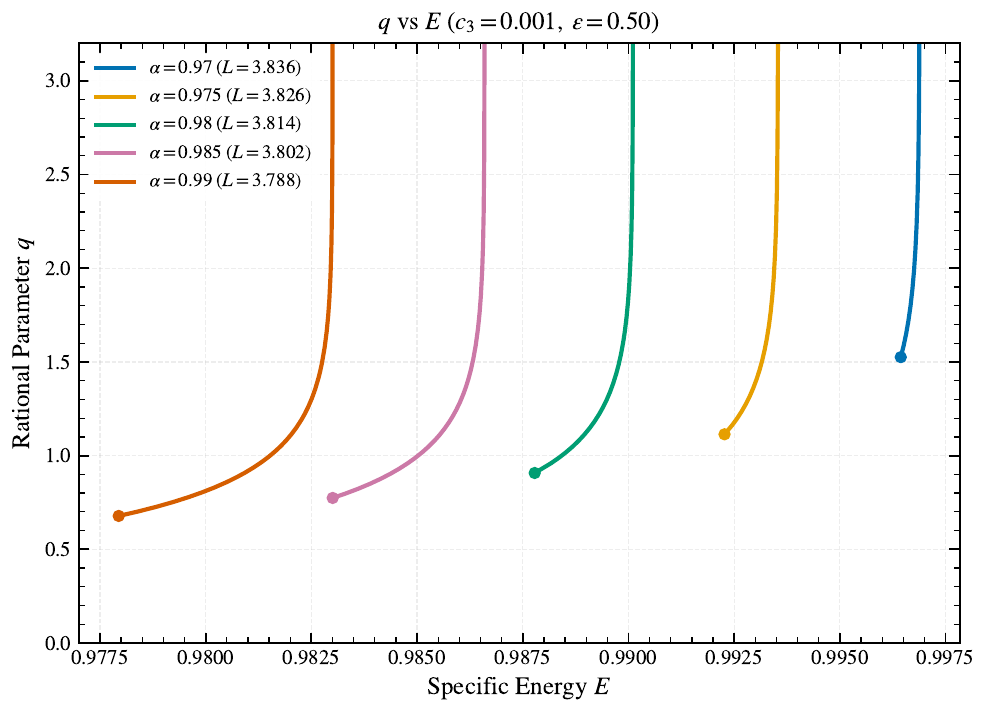}}
\caption{Rational parameter $q$ as a function of the particle energy $E$ for fixed angular momentum and varying $\alpha$.}
\label{fig06}
\end{figure}

\begin{figure}[htbp]
\centerline{
\includegraphics[scale = 0.45]{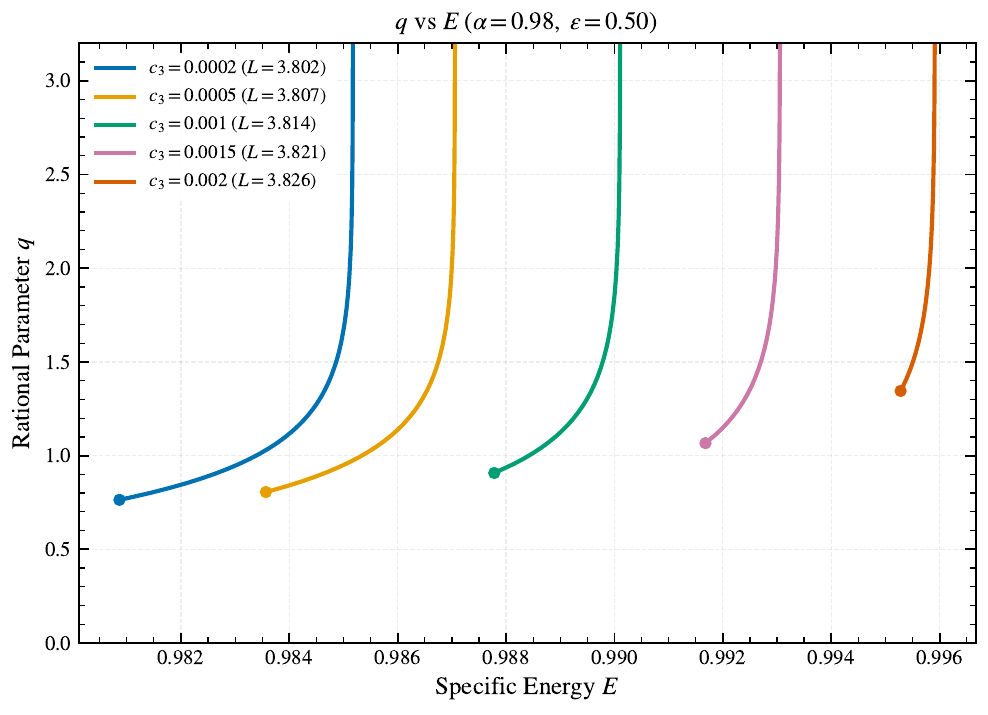}}
\caption{Rational parameter $q$ as a function of the particle energy $E$ for fixed angular momentum and varying $c_3$.}
\label{fig07}
\end{figure}

Figures~\ref{fig06} and \ref{fig07} illustrate the rational parameter $q$ as a function of the particle energy $E$ for several fixed values of the angular momentum. The parameter $q$ diverges as $E$ approaches its critical value associated with the unstable circular orbit, signaling the onset of extreme zoom--whirl behaviour. For larger $L$, the divergence occurs at higher energies, indicating that stronger angular momentum delays the transition to whirl-dominated motion. This demonstrates that the topological structure of periodic orbits is highly sensitive to the energetic proximity to the unstable orbit, and that the Weyl spacetime supports a rich hierarchy of zoom--whirl trajectories.

\begin{figure*}[htbp]
\centerline{
\includegraphics[scale = 0.58]{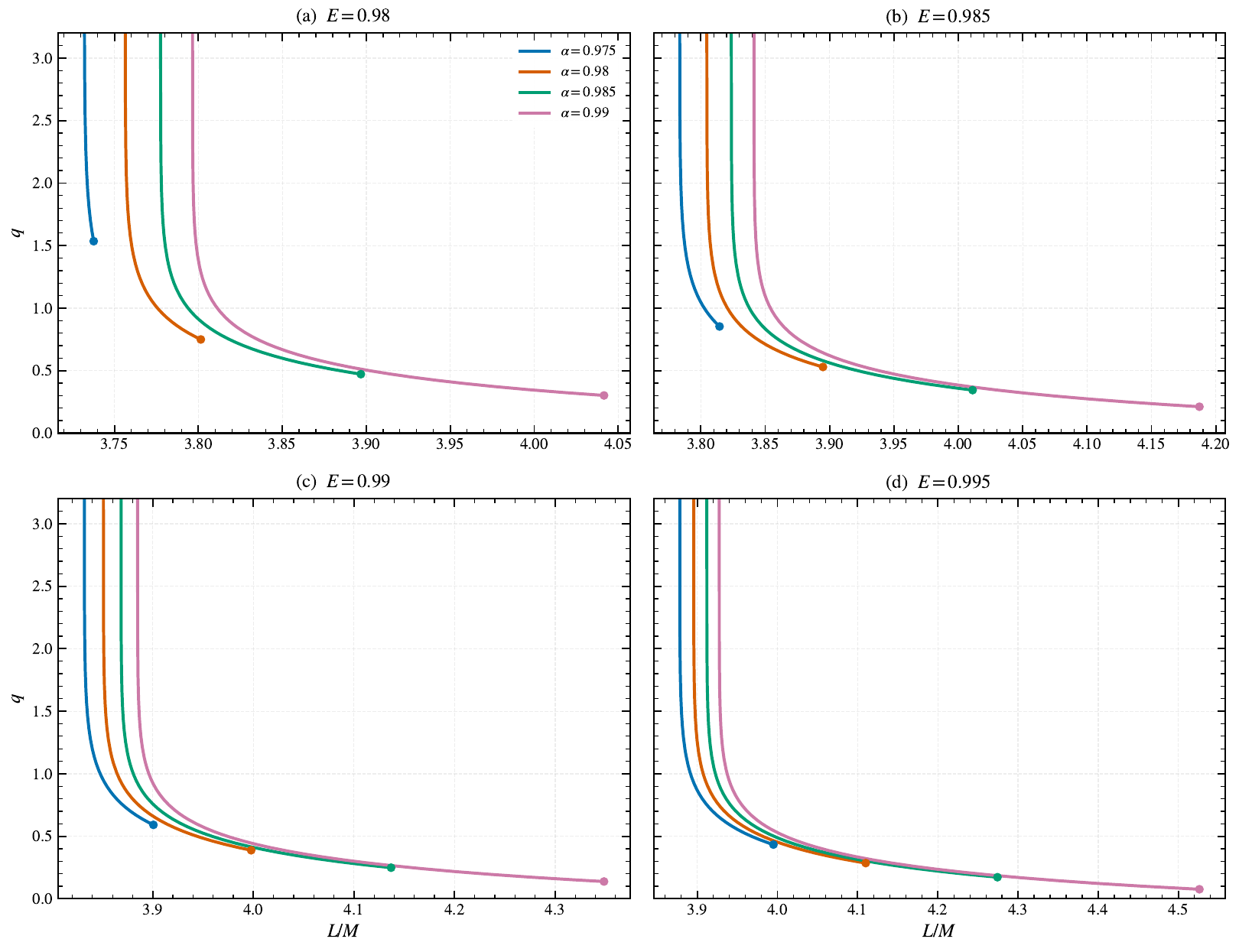}}
\caption{Rational parameter $q$ as a function of the angular momentum $L$ for fixed energy and varying $\alpha$.}
\label{fig08}
\end{figure*}

\begin{figure*}[htbp]
\centerline{
\includegraphics[scale = 0.58]{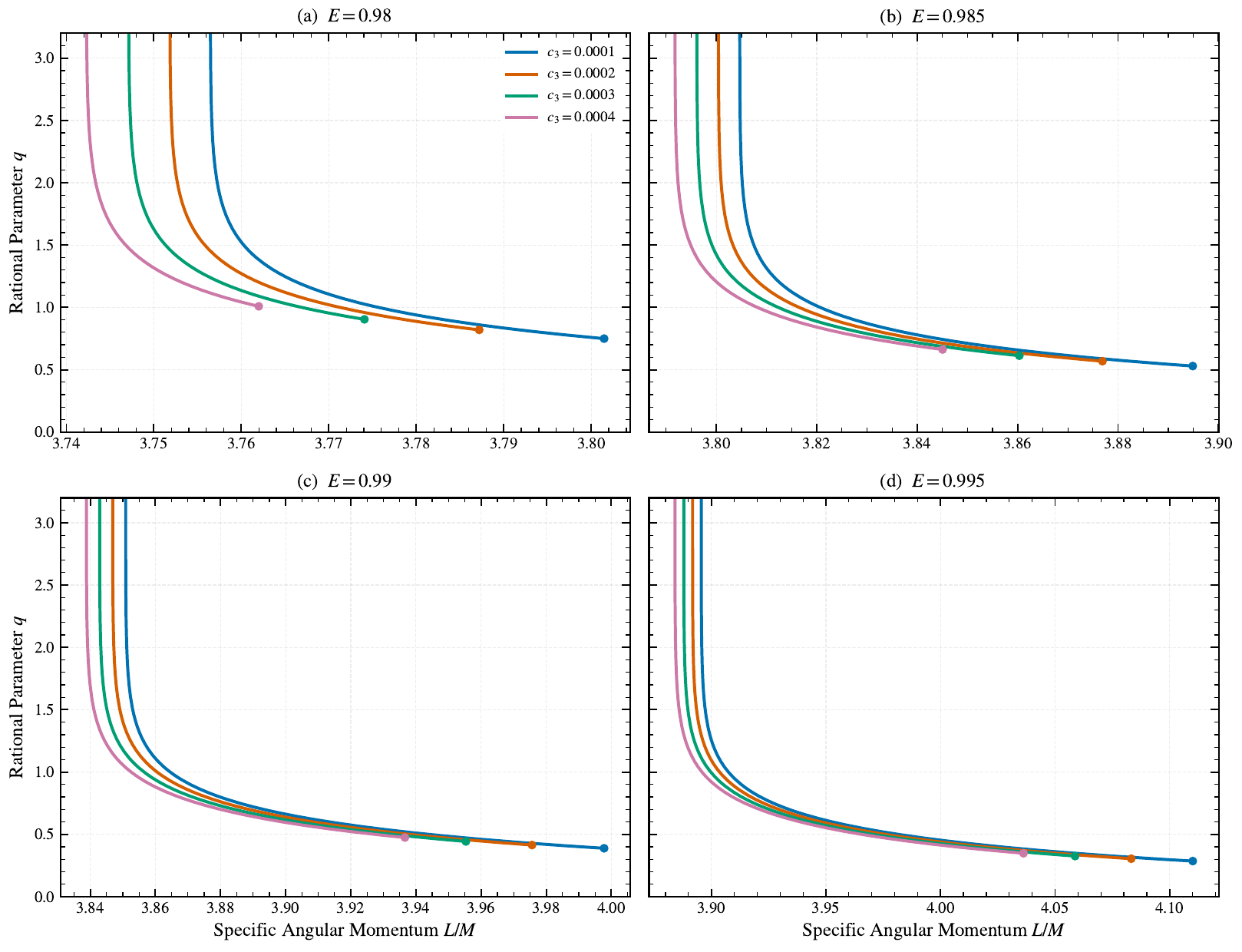}}
\caption{Rational parameter $q$ as a function of the angular momentum $L$ for fixed energy and varying $c_3$.}
\label{fig09}
\end{figure*}

Figures~\ref{fig08} and \ref{fig09} show the rational parameter $q$ as a function of the angular momentum $L$ for several fixed values of the particle energy $E$ in the Weyl spacetime. Each panel corresponds to a different energy level, while the coloured curves represent different values of the Weyl parameters ($\alpha$ and $c_3$). The rational parameter $q$ encodes the ratio between the azimuthal and radial frequencies and thus provides a topological classification of periodic orbits.

For fixed energy, $q$ decreases monotonically as the angular momentum increases. This behaviour reflects the fact that larger angular momentum suppresses whirl motion and promotes more regular, nearly Keplerian trajectories. In the limit of large $L$, the orbits become weakly relativistic, and the precession per radial cycle decreases, leading to smaller values of $q$.

In contrast, as $L$ approaches its critical lower bound associated with the unstable circular orbit, the rational parameter diverges, $q\rightarrow\infty$. This divergence signals the onset of extreme zoom--whirl behaviour, in which the particle executes an arbitrarily large number of revolutions near the unstable orbit before escaping back to the outer region. This regime corresponds to strong-field dynamics dominated by the near-horizon geometry.

The effect of the Weyl parameters is clearly visible in the horizontal shift of the curves. Modifying $\alpha$ and $c_3$ systematically moves the divergence point toward smaller or larger values of $L$, indicating that the geometric nonmetricity dictates the threshold for whirl-dominated dynamics. As the energy increases, the divergence points shift to larger values of $L$ and the overall magnitude of $q$ decreases. This implies that higher-energy particles require larger angular momentum to sustain bound whirl motion, making extreme zoom--whirl behaviour less pronounced at higher energies within the Weyl gravity framework.

\subsection{Topological Classification and Zoom-Whirl Orbits}\label{s4sub4}

Periodic orbits can be systematically classified using the topological indices $(z, w, v)$, which represent the zoom number, whirl number, and vertex number, respectively. The zoom number $z$ indicates the number of radial oscillations (outer lobes) the particle completes before the orbit closes, while the whirl number $w$ denotes the number of full azimuthal revolutions executed near the periastron during the strong-field phase. The vertex number $v$ characterizes the angular offset of the lobes. This classification provides a comprehensive topological mapping of relativistic periodic motion in the strong-field regime.

\begin{figure*}[htbp]
\centerline{
\includegraphics[scale = 0.58]{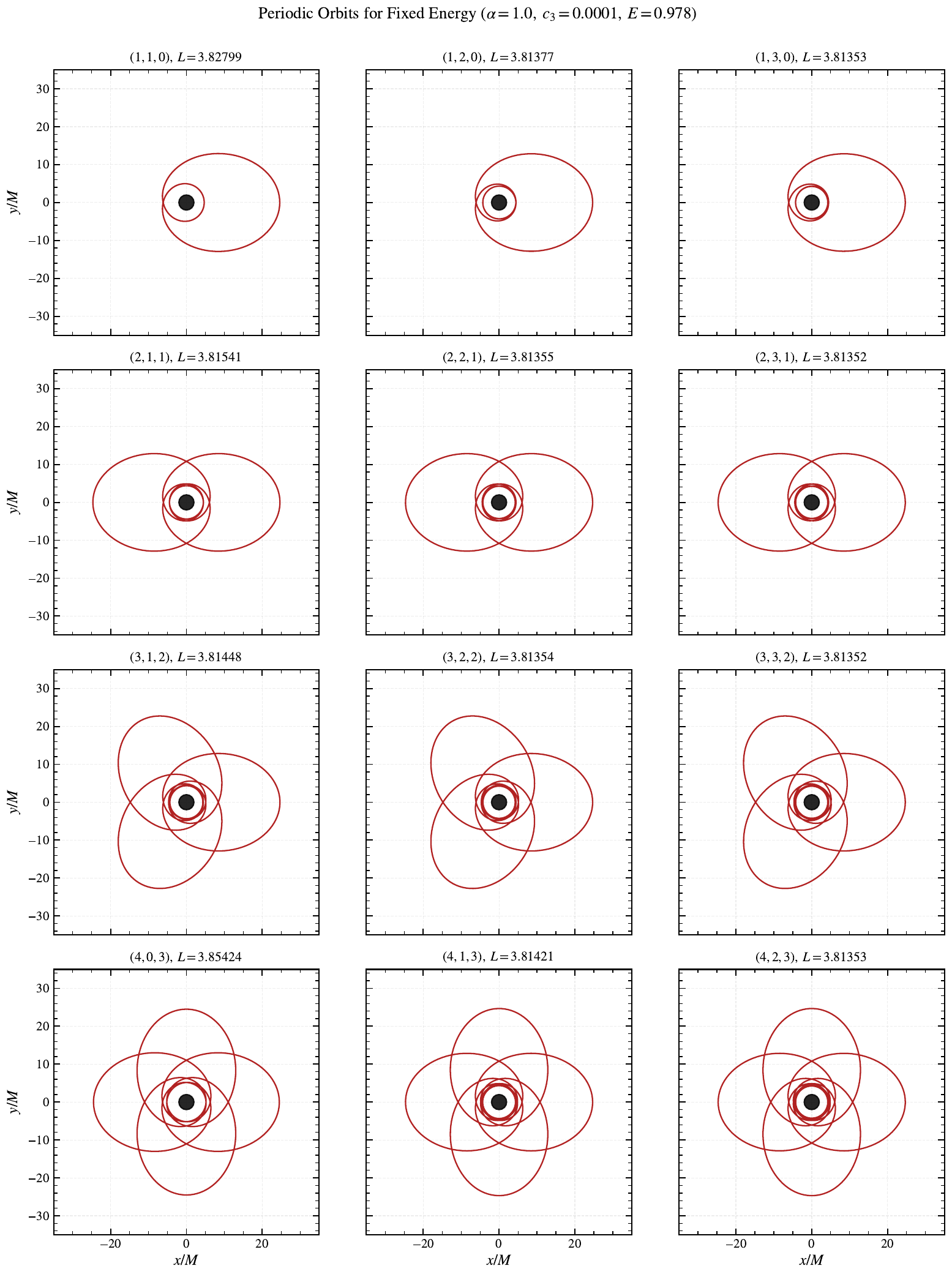}}
\caption{Periodic orbits in the Weyl geometry black hole for a fixed energy $(E = 0.978)$, characterized by different topological indices $(z, w, v)$. The parameters are set to $\alpha = 1.0$ and $c_3 = 0.0001$.}
\label{fig10}
\end{figure*}

\begin{figure*}[htbp]
\centerline{
\includegraphics[scale = 0.58]{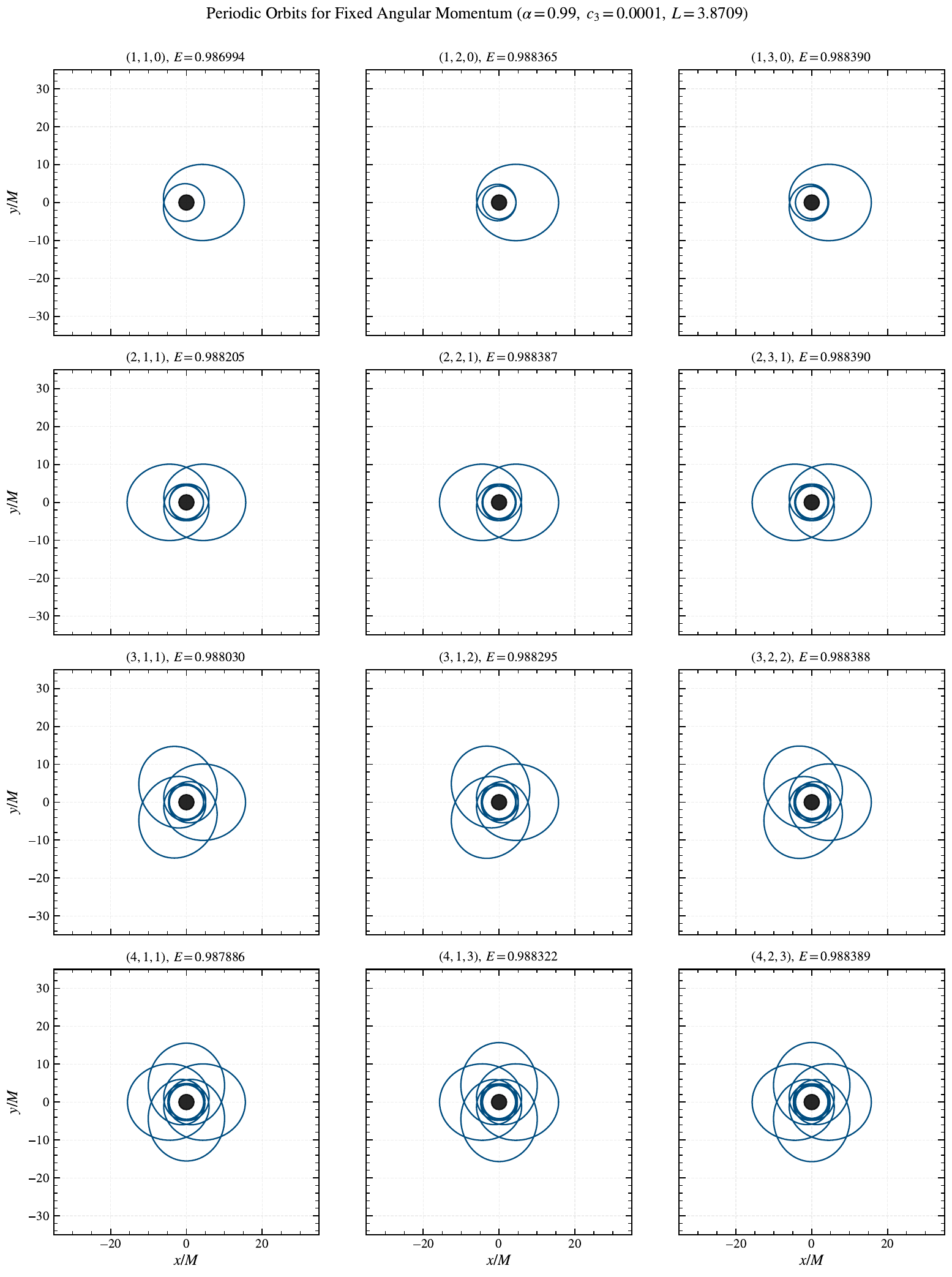}}
\caption{Periodic orbits in the Weyl geometry black hole for a fixed angular momentum $(L = 3.8709)$, characterized by different topological indices $(z, w, v)$. The parameters are set to $\alpha = 0.99$ and $c_3 = 0.0001$.}
\label{fig11}
\end{figure*}

Figures \ref{fig10} and \ref{fig11} display representative families of periodic equatorial orbits around the Weyl geometric black hole.

Figure \ref{fig10} illustrates orbits for a fixed specific energy $(E = 0.978)$ where the angular momentum $L$ is finely tuned to realize different rational $q$ values. As the zoom number $z$ increases (observed by moving down the rows), the trajectory develops additional outer leaves, indicating an increased number of radial excursions. Moving across the columns, larger values of the whirl number $w$ correspond to more tightly wound loops near the center of the potential well. This topology indicates prolonged trapping of the test particle close to the unstable circular orbit. Notably, a very minute decrease in $L$ pushes the periastron deeper into the strong gravitational field, drastically increasing the relativistic precession and leading to multiple inner whirls before the particle zooms back out.

Figure \ref{fig11} presents periodic orbits for a fixed angular momentum $(L = 3.8709)$ in a spacetime with active nonmetricity deformations ($\alpha = 0.99$ and $c_3 = 0.0001$). In this scenario, the specific energy $E$ is adjusted to achieve different topological classes. As $E$ is tuned closer to the critical maximum of the effective potential, the particle spends significantly more proper time near the barrier, resulting in higher whirl numbers.

Comparing Figure \ref{fig10} and Figure \ref{fig11} highlights the distinct roles of $E$ and $L$ in shaping the orbital dynamics. Varying the energy primarily controls the global size and radial excursion (the apastron distance) of the orbit. Conversely, varying the angular momentum heavily regulates the depth of strong-field trapping and the precise onset of whirl activity.

These plots confirm that the Weyl geometric black hole spacetime supports a highly intricate hierarchy of extreme zoom-whirl trajectories. While the fundamental topological behavior mirrors that of standard general relativity, the presence of the Weyl nonmetricity parameter $\alpha$ and the scaling parameter $c_3$ systematically shifts the exact energetic and angular momentum thresholds required to trigger these specific resonances. Such shifts have direct astrophysical implications, particularly regarding the phase evolution of gravitational waves emitted during extreme mass-ratio inspirals (EMRIs) in a Weyl geometric background.

\begin{figure*}[htbp]
\centerline{
\includegraphics[scale = 0.58]{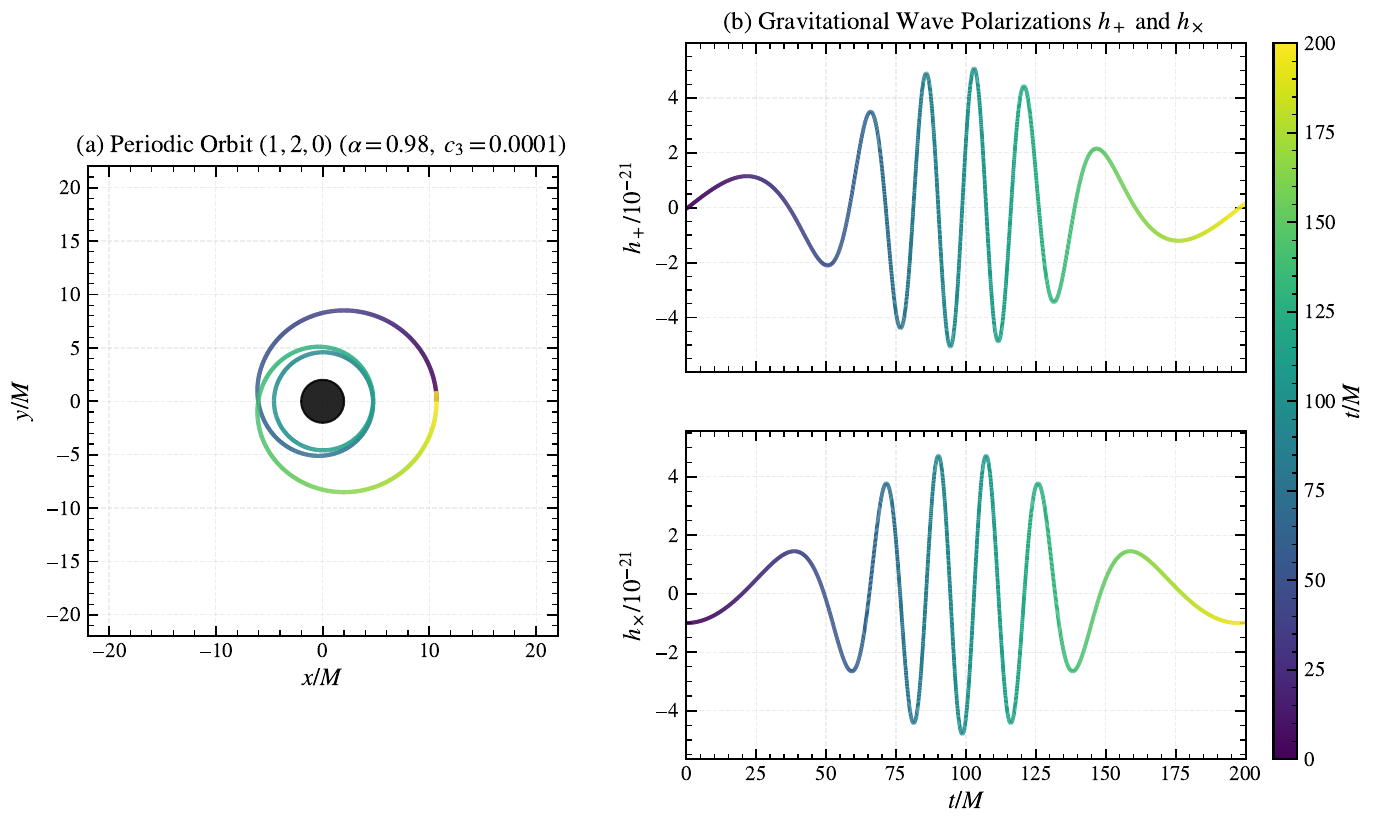}}
\caption{Gravitational waveform associated with the periodic orbit $(z,w,v) = (1, 2, 0)$ in the Weyl geometry black hole spacetime with parameters $\alpha=0.98$ and $c_3=0.0001$. The left panel shows the corresponding periodic orbit, while the right panels display the GW polarizations $h_+$ and $h_\times$.}
\label{fig12}
\end{figure*}

\begin{figure*}[htbp]
\centerline{
\includegraphics[scale = 0.58]{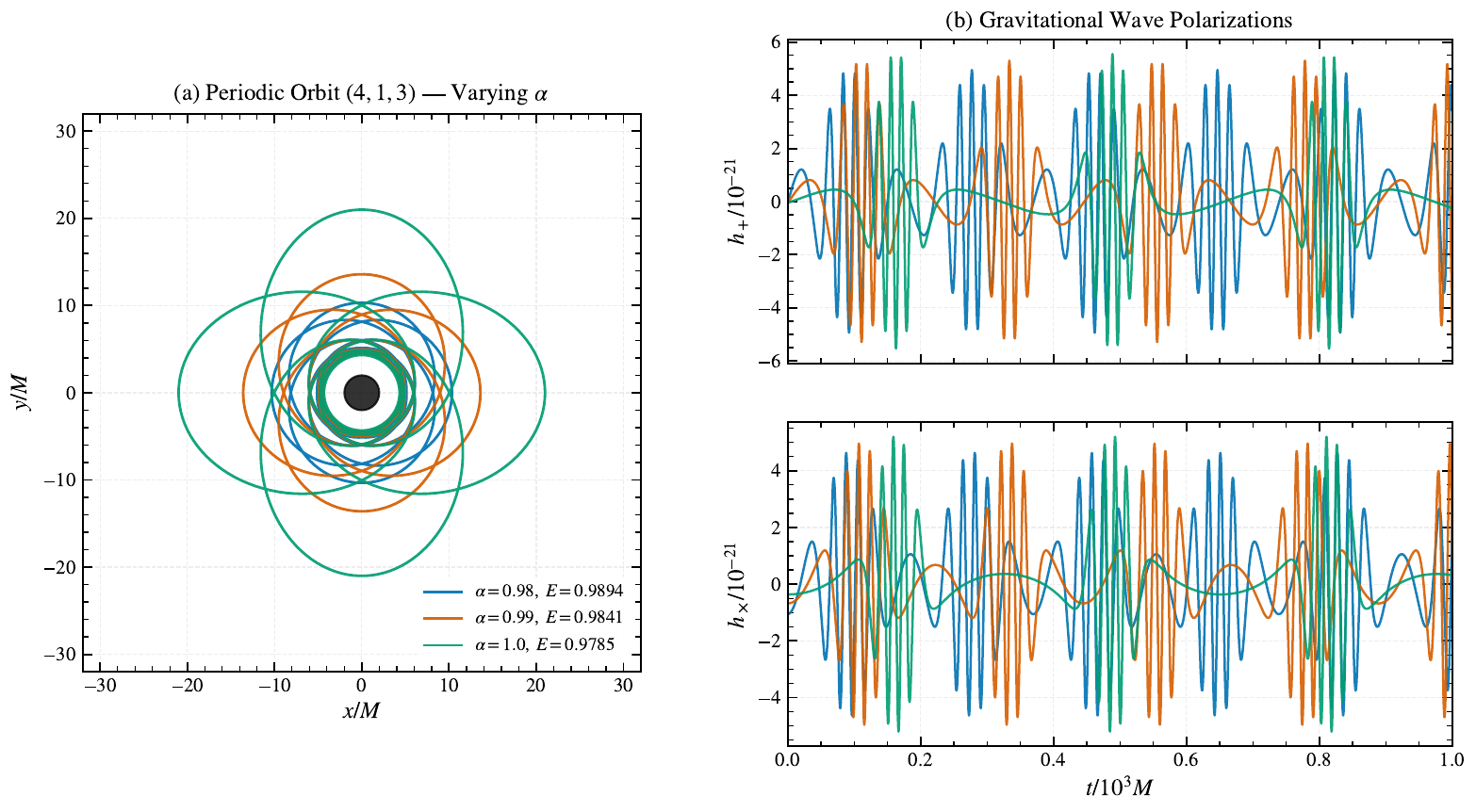}}
\caption{Effect of varying the Weyl nonmetricity parameter $\alpha$ on the orbital geometry and waveforms for the $(4,1,3)$ periodic orbit, with fixed $c_3 = 0.0002$ and $\epsilon = 0.65$. Increasing $\alpha$ expands the orbital lobes and induces noticeable phase shifts in the GW signal.}
\label{fig13}
\end{figure*}

\begin{figure*}[htbp]
\centerline{
\includegraphics[scale = 0.58]{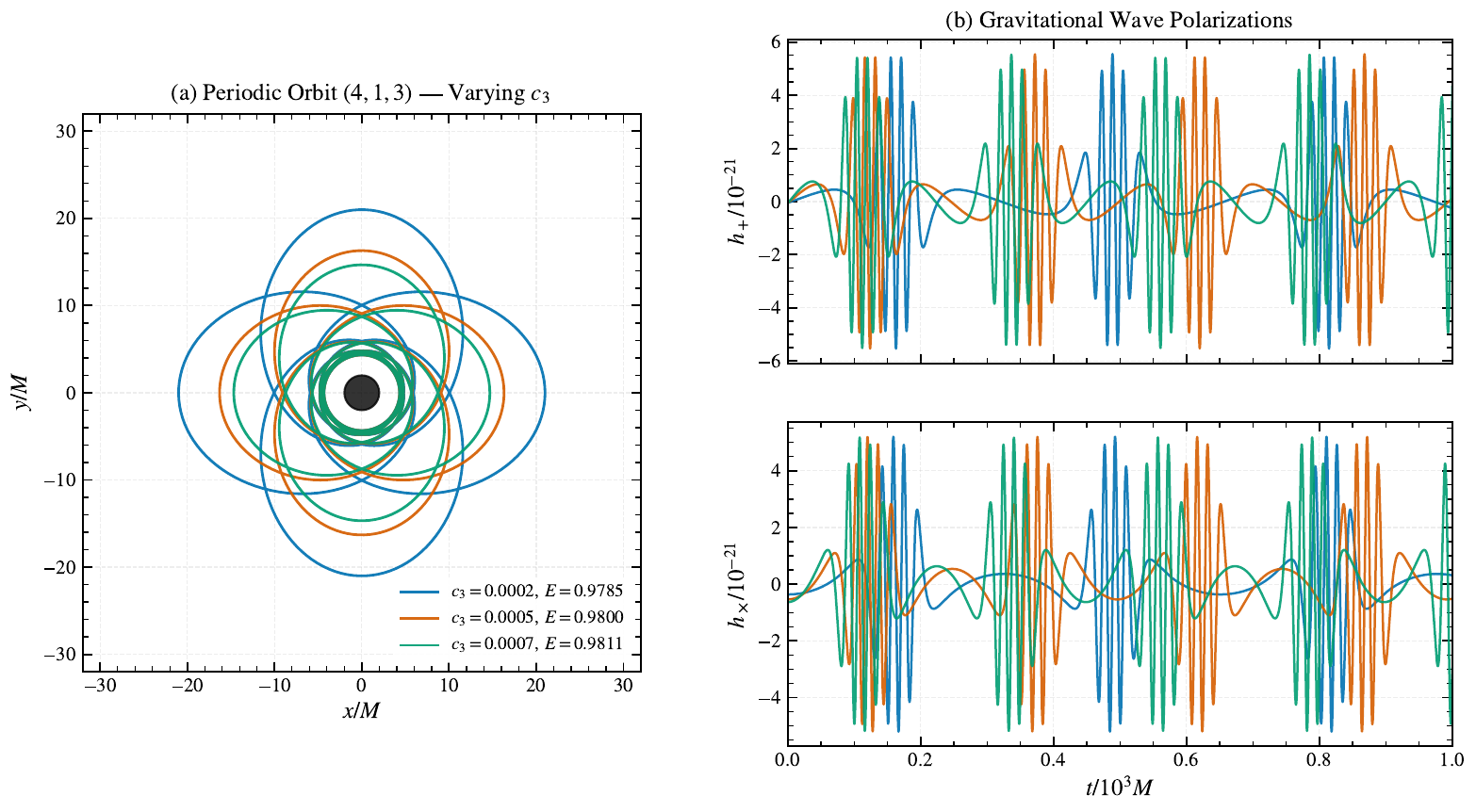}}
\caption{Effect of varying the parameter $c_3$ on the periodic orbit $(4,1,3)$ and its waveforms, with fixed $\alpha = 1.0$ and $\epsilon = 0.65$. Larger values of $c_3$ shrink the outer lobes, increasing the radial frequency and pulling the GW bursts closer together.}
\label{fig14}
\end{figure*}

\section{Numerical Kludge Gravitational Wave Radiation from Periodic Orbits}\label{sect3}

In an extreme mass-ratio inspiral (EMRI) system, the Weyl geometric black hole, described by the metric function $f(r)$ in Eq.~\eqref{ds}, can effectively model a central supermassive compact object. A small body moving along a periodic bound orbit in this strong-field spacetime acts as a localized source of gravitational radiation. Because the mass of the orbiting object is assumed to be much smaller than that of the central black hole ($m \ll M$), its backreaction on the background geometry can be neglected, and the particle's motion may be treated as a perturbation on a fixed background.

In this regime, the adiabatic approximation is highly applicable, which assumes that the loss of energy and angular momentum due to gravitational radiation is negligible over the course of a single orbital period \cite{Hughes:2000ssa, Glampedakis:2002ya, Zhang:2025wni}. Consequently, the constants of motion $(E,L)$ may be treated as fixed while computing the waveform associated with a distinct periodic orbit. This phenomenological approach has proven reliable for modelling EMRIs and capturing the essential strong-field orbital dynamics \cite{Zhang:2025wni, Li:2025eln, Zhao:2024exh,Meng:2024cnq}.

To generate the waveforms, we employ the numerical kludge method \cite{BabakEtAl:2007}. The procedure consists of two main steps: (i) the periodic timelike geodesics are obtained by numerically integrating the exact equations of motion in the Weyl geometric spacetime; (ii) the resulting strong-field trajectories are mapped into a pseudo-flat background and utilized as sources in the standard quadrupole formula to generate the gravitational waveform.

The geodesics are initially solved in spherical coordinates $(r,\theta,\phi)$. For waveform generation, the trajectory is projected into a Cartesian frame $(x,y,z)$ defined by \cite{Thorne:1980ru, Zhang:2025wni, BabakEtAl:2007}:
\begin{equation}
x=r\sin\theta\cos\phi,\quad
y=r\sin\theta\sin\phi,\quad
z=r\cos\theta.
\end{equation}

Within the linearized gravity approximation, the metric perturbation $h_{\mu\nu}$ on a flat background $\eta_{\mu\nu}$ satisfies $\square \bar{h}^{\mu\nu}=-16\pi T^{\mu\nu}$ under the Lorentz gauge. In the slow-motion limit, the gravitational waveform is well-approximated by the quadrupole formula:

\begin{equation}
\bar{h}_{ij}^{\rm TT}(t,\mathbf{x})
=
\frac{2}{D_L}
\ddot I_{ij}^{\rm TT}(t-D_L),
\end{equation}
where $D_L$ is the luminosity distance to the source, and the mass quadrupole moment is given by:
\begin{equation}
I^{ij}=\int x^i x^j T^{tt}(t,\mathbf{x})\, d^3x.
\end{equation}

For a point particle of mass $m$ following the trajectory $\mathbf{Z}(t)$, the stress-energy tensor component is $T^{tt}(t,\mathbf{x}) = m\,\delta^3(\mathbf{x}-\mathbf{Z}(t))$. This yields the spatial trace-reversed metric perturbation \cite{Zhao:2024exh,Meng:2024cnq}:
\begin{equation}
h_{ij}=\frac{2m}{D_L}
\left(a_i x_j + a_j x_i + 2 v_i v_j \right),
\end{equation}
where $v_i$ and $a_i$ represent the velocity and acceleration of the particle, respectively.

To observe the waveforms, the detector frame $(X,Y,Z)$ is established by rotating the source frame by an inclination angle $\iota$ and a longitude of pericenter $\zeta$:

\begin{eqnarray}
\hat{e}_X &=& (\cos\zeta,-\sin\zeta,0),\\
\hat{e}_Y &=& (\cos\iota\sin\zeta,\cos\iota\cos\zeta,-\sin\iota),\\
\hat{e}_Z &=& (\sin\iota\sin\zeta,\sin\iota\cos\zeta,\cos\iota).
\end{eqnarray}
The two physical gravitational wave polarizations are then extracted via\cite{Yang:2024lmj, Meng:2024cnq}:
\begin{eqnarray}
h_+ &=& \frac{1}{2}(e_X^i e_X^j - e_Y^i e_Y^j) h_{ij},\\
h_\times &=& \frac{1}{2}(e_X^i e_Y^j + e_Y^i e_X^j) h_{ij}.
\end{eqnarray}

For our numerical setup, we consider a representative EMRI with a compact object of mass $m=10M_\odot$ orbiting a supermassive Weyl black hole of mass $M=10^6M_\odot$. We assume an inclination $\iota=\pi/4$, $\zeta=\pi/4$, and a luminosity distance $D_L=200\,{\rm Mpc}$.

The emitted waveforms tightly encode the geometric features of the background spacetime, exhibiting a clear imprint of the underlying zoom--whirl orbital structure. Figure~\ref{fig12} illustrates a representative gravitational waveform associated with the $(z,w,v)=(1,2,0)$ periodic orbit for Weyl parameters $\alpha=0.98$ and $c_3=0.0001$. During the zoom phase, where the particle travels through the weaker gravitational field at larger radii, the radiation remains smooth with a relatively low amplitude. Conversely, during the whirl phase, the particle executes rapid, tight revolutions near the unstable circular orbit. The strong acceleration experienced in this high-curvature region produces sharp, high-frequency bursts in both GW polarizations $h_{+}$ and $h_{\times}$. The envelope of these peaks precisely corresponds to the brief intervals of strong-field trapping.

Figure~\ref{fig13} demonstrates the effect of varying the Weyl parameter $\alpha$ on a fixed topological class $(4,1,3)$. As $\alpha$ increases towards $1.0$, the overall size of the orbit expands, pushing the apastron outward. This geometric deformation directly alters the time the particle spends in the zoom phase, leading to significant phase shifts in the generated waveforms. Furthermore, the variation in $\alpha$ modifies the depth of the effective potential, which subtly alters the peak amplitudes of the radiation during the whirl phases.

The distinct influence of the parameter $c_3$ is depicted in Figure~\ref{fig14}. For a fixed $\alpha=1.0$, increasing $c_3$ has a contracting effect on the orbital geometry; the outer lobes shrink significantly, bringing the apastron closer to the black hole. This inward shift increases the overall radial frequency of the orbit. As a result, the time interval between successive whirl phases decreases, causing the high-amplitude gravitational wave bursts to occur closer together in time.

Collectively, these kludge waveforms illustrate that the nonmetricity of Weyl geometric gravity leaves distinct, measurable signatures in the EMRI gravitational wave signals. The specific values of $\alpha$ and $c_3$ jointly dictate the timing, amplitude, and frequency evolution of the zoom-whirl bursts, providing a potential observational avenue to test Weyl geometric gravity using future space-based gravitational wave detectors.

\section{Concluding Remarks}\label{sect4}

In this paper, we have systematically investigated the astrophysical properties of an exact, static, and spherically symmetric black hole solution derived within the framework of Weyl geometric gravity. Unlike standard General Relativity, this theory incorporates a nonmetric structure dictated by a Weyl vector, yielding a modified spacetime metric parameterized by the black hole mass $M$ and the Weyl geometric parameters $\alpha$ and $c_3$. By analyzing the timelike geodesics, we explored how this geometric nonmetricity influences the strong-field dynamics of massive test particles, the topological structure of periodic orbits, and the resulting gravitational wave emissions in extreme mass-ratio inspiral (EMRI) systems.

Our analysis of the effective potential revealed that the Weyl parameters introduce a controlled deformation of the gravitational potential well and barrier. We found that the locations and properties of the innermost stable circular orbit (ISCO) and the marginally bound orbit (MBO) are highly sensitive to these deformations. Specifically, an increase in $\alpha$ expands the ISCO radius while lowering the required specific energy and angular momentum for stable circular motion. Conversely, increasing the parameter $c_3$ shifts the stability threshold inward, requiring particles to possess higher energy and angular momentum to avoid plunging into the black hole. Consequently, the permissible phase space for long-lived bound orbits in the $(E, L)$ plane is either compressed or expanded depending on the specific values of the Weyl parameters.

To further characterize the strong-field dynamics, we examined the periodic bound trajectories by evaluating the rational parameter $q$, which dictates the transition into the extreme zoom-whirl regime. Using the topological indices $(z, w, v)$, we mapped an intricate hierarchy of resonant orbits. The results demonstrate that while the fundamental zoom-whirl topology mirrors that of standard Reissner-Nordström or Schwarzschild geometries, the nonmetricity of the Weyl spacetime systematically shifts the energetic and angular momentum thresholds required to trigger these specific whirl-dominated resonances.

Finally, we computed the gravitational radiation emitted by a test body tracing these periodic orbits using the numerical kludge method. The resulting EMRI waveforms clearly encode the geometric imprints of the Weyl background. During the tight whirl phases near the unstable circular orbit, the particle emits sharp, high-frequency bursts of gravitational waves. We demonstrated that increasing $\alpha$ expands the orbital lobes and induces noticeable phase shifts in the GW signal, while increasing $c_3$ contracts the outer lobes, raising the radial frequency and pulling the high-amplitude GW bursts closer together in time.

In conclusion, the presence of nonmetricity in Weyl geometric gravity leaves distinct and measurable signatures on both the orbital topology and the emitted gravitational waveforms of compact EMRI systems. These unique temporal and amplitude modulations provide a promising observational avenue. Future space-based gravitational wave observatories could potentially detect these characteristic phase shifts and burst frequencies, offering a novel empirical test to distinguish Weyl geometric gravity from standard General Relativity in the strong-field regime.


\section*{Declaration of competing interest}
The authors declare that they have no known competing financial interests or personal relationships that could have appeared to influence the work reported in this manuscript.

\section*{Data Availability Statement}
There are no new data associated with this article.

\bibliography{refs}

\end{document}